\documentclass[pre,aps,onecolumn,superscriptaddress,longbibliography]{revtex4-2}

\usepackage{amsmath,amssymb}
\usepackage{graphicx}
\usepackage{bm}        
\usepackage{hyperref}
 \usepackage[usenames,dvipsnames]{color}
\usepackage{soul}
\usepackage{lipsum}
\usepackage{graphicx}
\usepackage{subcaption}
\usepackage{setspace}
\usepackage{float}
\usepackage[utf8]{inputenc}
\usepackage{tikz}
\usepackage{amsmath,amssymb,amsfonts,graphics,dcolumn,bm,enumerate}
\usepackage{comment,natbib,appendix}
\usepackage{multirow,color}
\usepackage{chngpage}
\usepackage{afterpage}
\usepackage{xcolor}
\usepackage{amsthm}
\usepackage{hyperref}
\usepackage{epstopdf}
\usepackage{float}
\usepackage{soul}
\usepackage{booktabs}
\usepackage{array}              
\begin{document}

\title{Relations Among Different Inequality Measures in Complex Systems: From Kinetic Exchange to Earthquake Models}

\author{Shohini Sen}
\email{shohini.sen@students.iiserpune.ac.in}
\affiliation{Indian Institute of Science Education and Research, Pune-411008, India}

\author{Suchismita Banerjee}
\email{suchib@bose.res.in}
\thanks{Corresponding author}
\affiliation{S. N. Bose National Centre for Basic Sciences, Kolkata-700106, India}
\author{Bikas K Chakrabarti}
\email{bikask.chakrabarti@saha.ac.in}
\affiliation{Indian Statistical Institute, Kolkata-700108, India}
\affiliation{Saha Institute of Nuclear Physics, Kolkata-700064, India}

\begin{abstract}
We present a numerical study of several inequality measures across two kinetic wealth-exchange models with extreme inequality features (namely the Banerjee model, and the Chakraborti or Yard-Sale model) and two earthquake simulating models (namely the Chakrabarti-Stinchcombe two-fractal overlap model and the nonlinear dynamical Burridge-Knopoff model). 
For each model we compute numerically the Lorenz function for the respective models’ wealth, overlap magnitude or avalanche distributions. 
We then estimate the variations of Gini ($g$), Pietra ($p$) and Kolkata ($k$) indices in these models with systematic variations of saving propensity (for the two wealth-exchange models), with systematic variations of generation or block numbers (for the two earthquake simulating models). 
We find that for appropriate values of the respective model parameters, the inequality indices $g$ and $k$ in corresponding the distributions (of `wealth'  or `avalanche') show quantitatively similar behavior, namely  $g = k \simeq 0.86$, which was identified earlier to correspond to the precursor point of criticality in self-organized critical models ($k = 0.80$ corresponds to that for Pareto’s 80-20 law). 
The values of $p/(2k-1)$ in all these (wealth exchange and earthquake) models  remain a little above unity, as was predicted theoretically. 
These observations for the inequality indices $g$, $k$ and $p$ across the socio-economic and geophysical models indicate the presence of unifying subtle features in the statistics of such disparate dynamical systems.
\end{abstract}

\maketitle

\section{Introduction}\label{sec:1}

Inequality arises not only in socio-economic systems but also in a wide range of physical and geophysical processes. To compare these behaviors within a unified framework, we numerically quantify inequality in four classes of models: two kinetic wealth-exchange models, a synthetic Pareto distribution, and two earthquake-simulating models.

For the socio-economic domain, early models introduced by Dragulescu and Yakovenko~\cite{DY_2000} and Chakraborti and Chakrabarti \cite{CC_2000} demonstrated that simple stochastic exchange rules can generate realistic wealth distributions. 
Extensions involving fixed or distributed saving propensity~\cite{CCM_2004}, interaction heterogeneity, and asymmetric exchange have yielded richer phenomenology. 
The Chakraborti Model~\cite{C_Model}, also known as Yard-Sale model~\cite{YS}, later extended by Boghosian and collaborators~\cite{YS1}, captures multiplicative dynamics and extreme wealth condensation. 
The Banerjee Model~\cite{B_Model,B_Model1} represents another natural kinetic exchange framework, whose intrinsic trading dynamics lead to steady-state configurations characterized by extreme concentration of wealth among a small subset of agents.
In this study, we focus on the Banerjee Model, where inequality is tuned by varying the range of exchange and the saving propensity, and the Chakraborti Model (or Yard-Sale model), where saving propensity acts as the central control parameter. 


We also examine the inequality associated with Pareto-type distributions~\cite{Pareto,Pareto1} by tuning their shape parameter, given the ubiquity of power laws in systems exhibiting self-organized criticality. Since such distributions are also central to geophysical phenomena, we analyze inequality patterns in two different earthquake models, the Chakrabarti-Stinchcombe two-fractal overlap model~\cite{CS_Model,CS1}, where event sizes arise from Cantor-set overlaps, and the nonlinear Burridge-Knopoff spring-block model \cite{langer_1994,BK_Model}, where block–spring dynamics generate heterogeneous slip events. The aim of this study is to explore the universal features of inequality measures (for the distribution of `wealth' or `avalanche') across these different complex socio-economic and geophysical models.

We first quantify the inequality across these systems by investigating three Lorenz curve~\cite{lor,lor1} based measures: the Gini index ($g$)~\cite{gini}, captures the average deviation from perfect equality; The Pietra (or Hoover) index ($p$)~\cite{pietra,pietra1}, measures the maximum imbalance in cumulative wealth and the more recently introduced Kolkata index ($k$)~\cite{k-index,k-index1,k-index2,k-index3}, identifies the wealth share of the richest fraction of the population and has found applications in diverse socio-economic contexts. 
For each system, we systematically vary the relevant control parameters such as the exchange range ($R$) and saving propensity ($\lambda$) in the Banerjee Model, saving propensity ($\lambda$) in the Chakraborti Model, the exponent ($\alpha$) in the synthetic Pareto distribution, the Cantor generation index in Chakrabarti-Stinchcombe two-fractal overlap model and the time evolution of the Burridge-Knopoff model, and compute the corresponding inequality indices.
In a recent work~\cite{asim_2025} authors studied the ratios $p/(2k-1)$  and  $p/g$ in some financial systems. 
We further examine the ratios $p/(2k-1)$, $p/g$ for these kinetic exchange models, the two earthquake simulating models, and the Pareto distribution to investigate  the variation of these and other relationships among the inequality measures $g$, $p$ and $k$.

The motivation for considering such a broad class of models is to test whether the Lorenz-based inequality measures $g$, $p$, and $k$ capture only domain-specific properties or whether they provide a common statistical framework across systems with very different microscopic dynamics. Wealth-exchange models and earthquake-generating models arise in completely different physical contexts, yet both produce highly heterogeneous distributions with dominant tails. This makes them natural candidates for a unified comparative study. In particular, while these indices quantify wealth concentration in socio-economic systems, they can also characterize the concentration of overlaps or avalanche sizes in geophysical systems, thereby linking inequality growth and the approach to extreme events within the same quantitative language.
Together, these analyses provide a unified quantitative perspective on how inequality evolves across these models that are governed by very different microscopic rules. We can see how tuning a small number of parameters systematically shapes these outcomes. The paper is organized as follows: Section~\ref{sec:2} introduces the Lorenz curves and the inequality measures investigated in this study. Section~\ref{sec:3} outlines the four models and the corresponding simulation framework. Section~\ref{sec:4} presents the numerical results and analysis. Finally, Section~\ref{sec:5} offers a brief summary of the main results obtained and the concluding remarks.

\section{Inequality Measures}\label{sec:2}

Quantifying inequality in any distribution whether describing wealth, event sizes, or released energy (hereafter called wealth) requires statistical measures that capture heterogeneity in a systematic manner. 
In this study, we numerically examine three statistical measures based on the Lorenz function~\cite{lor,lor1}, namely the Gini index~\cite{gini}, the Pietra (or Hoover) index~\cite{pietra,pietra1}, and the Kolkata index~\cite{k-index,k-index2}. 

The Lorenz function (or Lorenz curve) $L(x)$ is a foundational tool for visualizing and quantifying inequality. From this function many inequality measures are derived. 
For a non–negative wealth $y$ with probability density function $P(y)$, let $F(y)= \int_{0}^{y}P(t)dt$ denote the cumulative distribution function. The Lorenz function is then defined as:
\begin{equation}\label{eqn:lor}
L(x) = \frac{\displaystyle \int_{0}^{y(x)} t\,P(t)\,dt}
{\displaystyle \int_{0}^{\infty} t\,P(t)\,dt},
\qquad 0 \le x \le 1,
\end{equation}
where $y(x)$ is defined implicitly by $F(y(x))=x$.
Here, $x$ denotes the bottom $x$ fraction of the population (ranked by wealth), and $L(x)$ gives the corresponding cumulative fraction of total wealth they hold.
For a perfectly equal distribution, $L(x)=x$, producing a 45-degree line known as the perfect equality line (see in Fig.~\ref{fig:ineq}).  
Deviations of $L(x)$ below this line indicate inequality.
In the limiting case where all wealth is concentrated in a single individual, we get the absolute inequality line as seen in Fig.~\ref{fig:ineq}.


\subsection{\textbf{Gini Index}}\label{sec:gini}
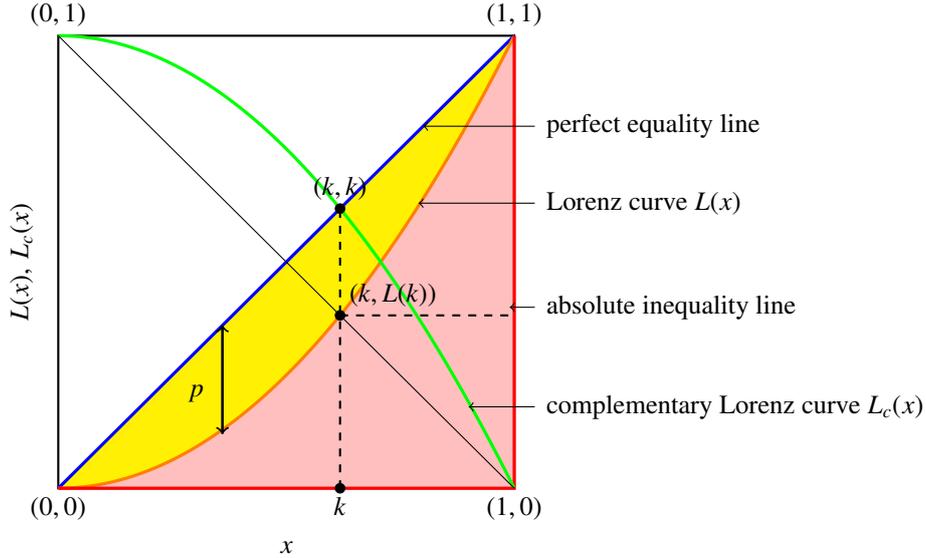
\begin{figure*}[t]
\centering
\begin{tikzpicture}[scale=5]
\def\xp{0.618}                        
\def\yp{0.382}
\draw[thick] (0,0) rectangle (1,1);

\draw[fill=yellow] (0,0) -- plot[domain=0:1,samples=200,smooth] (\x,\x*\x) -- (1,1);
\draw[fill=pink] plot[domain=0:1,samples=200,smooth] (\x,\x*\x) -- (1,1) -- (1,0) -- (0,0);

\draw[very thick,blue] (0,0) -- (1,1);

\draw[very thick,orange]plot[domain=0:1,samples=200,smooth] (\x,\x*\x);

 \draw[very thick,red] (0,0) -- (1,0) -- (1,1);

\draw[very thick,green] plot[domain=0:1,samples=200,smooth] (\x,{(1-\x*\x)});

\draw[thin] (0,1) -- (1,0);

\coordinate (Q) at (\xp,0);
\coordinate (R) at (\xp,\xp);
\coordinate (O) at (0.5,0.5);
\coordinate (P) at (\xp,\yp);

\draw[thick][dashed] (Q) -- (\xp,\xp); 
\draw[thick][dashed] (P) -- (1,\yp);

\node[above right]       at (P) {$(k,L(k))$};
\node[below]       at (Q) {$k$};
\node[above]       at (R) {$(k,k)$};
\filldraw [black] (\xp,\yp) circle (0.3pt);
\filldraw [black] (\xp,\xp) circle (0.3pt);
\filldraw [black] (\xp,0) circle (0.3pt);
\node[above,black]  at (0,1) {$(0,1)$};
\node[below,black]  at (0,0) {$(0,0)$};
\node[above,black] at (1,1) {$(1,1)$};
\node[below,black] at (1,0) {$(1,0)$};
\node[left,black] at (0.34,0.21) {$p$};
\draw[thick][->] (0.36,0.12) -- (0.36,0.36);
\draw[thick][->] (0.36,0.36) -- (0.36,0.12);
\node[very thick][rotate=90] at (-0.08,0.5){$L(x)$,\,\,$L_{c}(x)$};

\node [very thick] at (0.5,-0.13){$x$};

\node[right] at (1.05,0.80) {perfect equality line};
\draw[->] (1.045,0.80) -- (0.80,0.80);

\node[right] at (1.05,0.63) {Lorenz curve $L(x)$};
\draw[->] (1.045,0.63) -- (0.79,0.63);

\node[right] at (1.05,0.40) {absolute inequality line};
\draw[->] (1.045,0.40) -- (0.99,0.40);

\node[right] at (1.05,0.18) {complementary Lorenz curve $L_{c}(x)$};
\draw[->] (1.045,0.18) -- (0.90,0.18);
\end{tikzpicture}
\caption{Schematic diagram of Lorenz curve and inequality measures where x-axis represents cumulative fraction of population from poor to rich and y-axis depicts the cumulative fraction of their wealth. Gini index $g$ is given by the ratio of the area of yellow region and yellow$+$pink region. Pietra index $p$ is the maximum distance between the equality line and Lorenz curve and Kolkata index is the ordinate point $k$ where Lorenz curve cuts the off diagonal line (perpendicular to the equality line) at the point $(k,L(k))$.}
\label{fig:ineq}
\end{figure*}

The Gini index $g$ is one of the most widely used scalar measures of inequality, derived directly from the Lorenz curve. 
It ranges from $0$ to $1$. 
From Fig.~\ref{fig:ineq}, one can see that $g$ measures the ratio of the area between the equality (blue) line and the Lorenz curve L(x) (orange curve) and the area between the equality (blue) line and the absolute inequality (red) line.
In terms of the Lorenz curve:
\begin{equation}\label{eqn:gini}
    g = \frac{\int_{0}^{1}[x-L(x)]\,dx}{1/2}=1 - 2\int_{0}^{1} L(x)\,dx.
\end{equation}

\subsection{\textbf{Pietra Index}}\label{sec:pietra}
The Pietra index $p$ (also called the Hoover index or the Robin Hood index) measures the maximum vertical distance between the Lorenz curve and perfect equality line (see Fig.~\ref{fig:ineq}). 
Equivalently, it represents the excess fraction of the wealth that must be redistributed from the richer people to the poorer people to achieve perfect equality. 
The Pietra index takes values in $[0,1]$, with higher values implying stronger inequality. 
It is defined as:
\begin{equation}\label{eqn:pietra}
    p = \max_{0 \le x \le 1} \left[\, x- L(x)\,\right].
\end{equation} 

\subsection{\textbf{Kolkata Index}}\label{sec:k-index}
The Kolkata index (or $k$-index) is a more recent inequality measure, widely used in socio-economic systems and increasingly in complex physical systems.  
It is defined as the fixed point of the complementary Lorenz function ($L_{c}(x)$):
\begin{equation}\label{eqn:k-index}
    L_{c}(k)\equiv 1 - L(k) = k.
\end{equation}
In Fig.~\ref{fig:ineq}, from simple geometry one can see that $k$ is the fraction of the total wealth possessed by the richest $(1-k)$ fraction of the population.  
Note that, $k=0.80$ implies that $20\%$ of the population holds $80\%$ of the wealth, reminiscent of the Pareto $80$–$20$ rule.  
The $k$-index varies from $1/2$ (perfect equality) to $1$ (perfect inequality).  

In many empirical systems and theoretical models, $k$ displays near-universal scaling behaviour in relation to $g$ or $p$, which motivates our numerically comparative study across the two kinetic wealth exchange models and the two earthquake simulating models~\cite{sand,manna,soumya,asim_2025}.

\section{Models}\label{sec:3}

In this section, we describe the four models 
to investigate inequality across these socio-economic and geophysical systems. 
Our analysis spans two kinetic wealth-exchange models, a synthetic Pareto distribution, and two earthquake-simulating models based on fractal overlap and spring–block dynamics. 

\subsection{\textbf{Kinetic Wealth-Exchange Models}}\label{sec:KWE}

Kinetic wealth exchange models given by Dragulescu and Yakovenko~\cite{DY_2000} describe wealth evolution through pairwise stochastic exchanges among agents and those given by Chakraborti and Chakrabarti, through saving propensities~\cite{CC_2000}.
Below we describe the two kinetic wealth exchange models.


\subsubsection{\textbf{B-Model with Exchange Range}}\label{sec:B-Model}

The Banerjee model~\cite{B_Model,B_Model1} (or B-Model in short hereafter) introduces a constrained exchange rule in which each agent interacts only within a selected exchange range $R$ in the wealth-ordered list. At each time step $t$:\\
(i) Arrange agents by their current wealth from poor to rich.\\
(ii) Choose an agent $i$ at random.\\
(iii) Select the interaction or exchange partner $j$ randomly among the agents within range $R$ of $i$ in the wealth-ordered list.\\
The update rule reads:
\begin{equation}\label{eqn:B-Model}
w_i(t+1)= \epsilon (w_i(t) + w_j(t)), \qquad 
w_j(t+1)= (1-\epsilon) (w_i(t) + w_j(t)),
\end{equation}
where $\epsilon$ is a random number drawn uniformly from $(0,1)$.  
This model naturally generates strong wealth condensation for small $R$, producing highly unequal steady states, though not to the extreme limit observed in the C-Model (discussed later), where all wealth collapses into a single agent. 
For each choice of $R$, we compute the steady-state Lorenz geometry and associated inequality indices after ensemble averaging.

\subsubsection{\textbf{B-Model with Savings}}\label{sec:B-Model_with_l}

We also study a modified version of the Banerjee model in which agents retain a fixed fraction of their wealth before participating in the exchange. Each agent has a saving propensity $\lambda\in[0,1)$, so that only the remaining $(1-\lambda)$ portion of wealth is available for trading. The update rule becomes:
\begin{align}\label{eqn:B_with_l}
w_i(t+1)=\lambda w_i(t) + \epsilon(1-\lambda)(w_i(t) + w_j(t)), \nonumber\\
w_j(t+1)=\lambda w_j(t) + (1-\epsilon)(1-\lambda)(w_i(t) + w_j(t)),
\end{align}
where $\epsilon\in(0,1)$ is a random sharing parameter.  
The introduction of saving stabilizes the dynamics and suppresses extreme condensation, producing a spectrum of inequality levels as $\lambda$ is varied. For each value of $\lambda$, ensemble-averaged steady-state distributions are used to compute the inequality measures.

\subsubsection{\textbf{C-Model with Savings}}\label{sec:C-Model_with_l}

The Chakraborti Model~\cite{C_Model} (or C-Model in short hereafter) also widely known as Yard-Sale model~\cite{YS} is based on multiplicative dynamics in which the amount at stake depends on the wealth of the poorer agent. With saving propensity $\lambda \in [0,1)$, the effective tradable wealth becomes $(1-\lambda)\min\{w_i,w_j\}$. Therefore the exchanged amount is $\Delta w = \epsilon (1-\lambda)\min\{w_i(t), w_j(t)\}$, where $\epsilon$ is uniformly drawn from $(0,1)$.  
The update equations follow:
\begin{equation}\label{eqn:C_with_l}
w_i(t+1)=w_i(t) + \Delta w, \qquad 
w_j(t+1)=w_j(t) - \Delta w,
\end{equation}
with the sign depending on the random outcome of the exchange.  
This model is known for producing rich-get-richer dynamics and, for low saving propensities, may lead to complete condensation, where all wealth eventually accumulates with a single agent. Ensemble-averaged steady-state distributions are used for computing all inequality metrics.

We have discussed about two kinetic wealth-exchange models, both of which generate exponential or gamma-like distributions. 
We now turn to earthquake simulation models, which yield power-law distributions. Our focus here is not on analyzing the distributions produced by these models but rather on investigating inequality within them.
Before jumping from kinetic wealth exchange models to earthquake simulating models, we examine a synthetic Pareto distribution that exhibits different power-law behaviors as the exponent parameter ($\alpha$) is varied.

\subsection{\textbf{P-Model}}\label{sec:pareto}

We investigate inequality in a synthetic Pareto-type distribution~\cite{Pareto,Pareto1} (or P-Model in short hereafter) characterized by the probability density function,
\begin{equation}\label{eqn:Pareto}
P(x) = \frac{\alpha x_m^\alpha}{x^{\alpha+1}}, \qquad x \ge x_m,
\end{equation}
where $x_m$ is the minimum wealth (set to unity in our simulations) and $\alpha>0$ is the shape parameter.  
The parameter $\alpha$ controls tail heaviness: a smaller $\alpha$ produces broader, more unequal distributions. 
The Lorenz curve takes the form,
\begin{equation}\label{eqn:lor_p}
L(p) = 1 - (1 - p)^{\frac{\alpha - 1}{\alpha}},
\end{equation} 
where $p$ denotes the fraction of the population, and $L(p)$ is the corresponding cumulative fraction of total wealth, when the individuals are ordered in an ascending order of wealth.
For each $\alpha$ we generate large synthetic datasets, compute the Lorenz curve, and evaluate the associated inequality indices, averaging over ensembles. 
This baseline serves as a reference for comparing model-generated distributions with known power-law behavior observed in socio-economic and geophysical systems.

\subsection{\textbf{Earthquake Simulating Models}}\label{sec:EQS}

We now move to two classes of geophysical models that simulate earthquake-like event size statistics. 
The Chakrabarti-Stinchcombe fractal overlap construction generates heavy-tailed distributions that produce inequality through time variation of the overlap magnitude which arises due to unevenness of the surface. The Burridge-Knopoff model gives rise to inequality in terms of spatially varying spring force.
Both models probe different but complementary aspects of earthquake dynamics, namely, the geometric overlap of fractal faults, and stress distribution during an earthquake under non-linear friction. The Chakrabarti-Stinchcombe model exhibits a power-law scaling in the cumulative distribution of overlap magnitudes in the asymptotic limit of large generation number and large overlaps, arising purely from the construction of the underlying fractals~\cite{pbhattacharya}.
In contrast, the Burridge-Knopoff model displays power-law event-size statistics in specific dynamical regimes, where nonlinear friction and collective interactions drive the system toward scale-free behaviour.
Thus, although both models can produce similar phenomenological
power-law statistics, the underlying mechanisms are fundamentally different: geometric hierarchy in the former and dynamical evolution in the latter.

\subsubsection{\textbf{CS-Model}}\label{sec:CS-Model}

The Chakrabarti-Stinchcombe model~\cite{CS_Model,CS1,pbhattacharya} (CS-Model in short hereafter) based on the overlap of fractal fault surfaces (modeled using Cantor sets), focuses on fault geometry and the distribution of overlaps.
In its simplest realization, a Cantor set of any generation (representing a fractured part of the tectonic plate) slides uniformly in time over its replica (representing the rough continental lithosphere, having identical fractal surface structure) and the measure of the overlapping continuous sets of numbers (between  the two Cantor sets of identical generation, as one moves over the other) give the burst (or avalanche) sizes of elastic energies, modeling the continental stick-slip avalanches. It is therefore a simple geometric model of plate tectonics. It may explain why slip events follow broad distributions (e.g., power laws) from the self-similar structure of the fault interface.

\begin{figure}
    \centering   \includegraphics[width=0.43\linewidth]{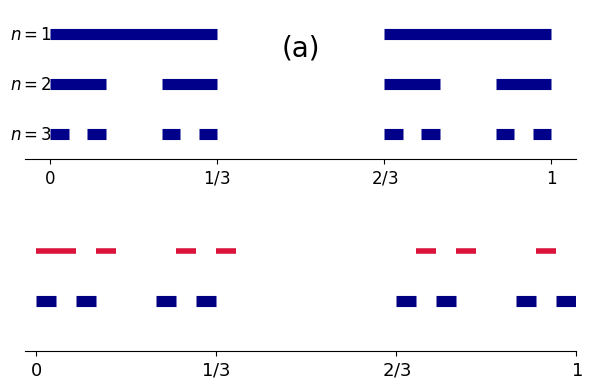}   \includegraphics[width=0.35\linewidth]{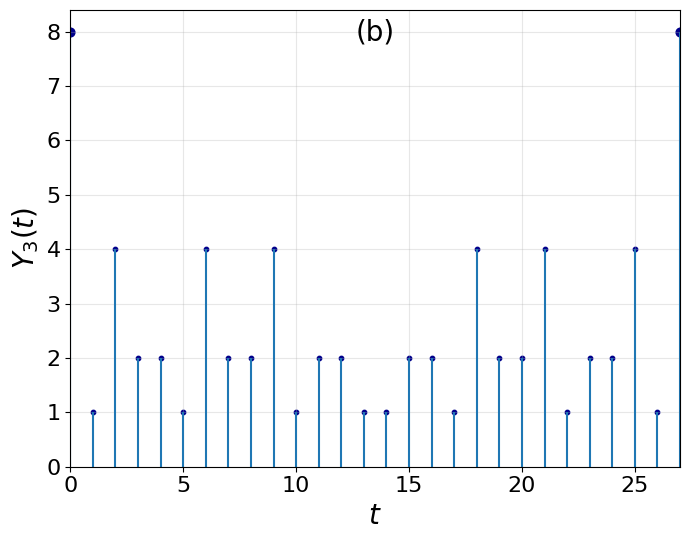}   
    \caption{(a) Construction of Cantor sets at generations $n = 1, 2, 3$ and overlaps $Y_3$ for two $n=3$ Cantor sets (with periodic boundary condition), as one set moves over the other. (b) Time series of overlap \(Y_3 (t)\)  as two Cantor sets of $n$ =3 slide over each other.}
    \label{fig:CS_cantor}
\end{figure}

The model treats a fault as two identical self-affine fractals, typically $n^{th}$-generation Cantor sets with $2^n$ segments of length $3^{-n}$, sliding over each other. 
A replica is shifted in discrete steps of $3^{-n}$ under periodic boundary conditions, and the overlap magnitude $Y_{n}(t)$ counts coinciding segments at step $t$, generating a discrete time series $\{Y_{n}(t)\}$. 
The allowed values of $Y_n$ are $2^{\,n-k}$ ($k=0,1,\dots,n$), occurring with exact frequencies \begin{equation}
N_r\!\left(2^{\,n-k}\right) = 2^k \binom{n}{k}
\label{eqn:CS Model}
\end{equation}
which normalize to the binomial probabilities $\Pr(2^{\,n-k}) = \binom{n}{k}\left(\tfrac{1}{3}\right)^{n-k}\left(\tfrac{2}{3}\right)^k$; the most probable overlap scales as $\hat{Y}_n \sim (2^n)^{1/3}$.
CS-Model captures the static, structural statistics of fault roughness providing insight into how geometric irregularity shape the statistics of earthquakes (see Fig.~\ref{fig:CS_cantor}).

\subsubsection{\textbf{BK-Model}}\label{sec:BK-Model}

In contrast, the Burridge-Knopoff Model~\cite{langer_1994,BK_Model} (BK-Model in short hereafter) is a dynamical and mechanical model of earthquake generation, where blocks connected by springs and subject to friction evolve over time. It is used to study stick–slip dynamics and stress redistribution. 
BK-Model captures the temporal dynamics of rupture propagation and seismicity and provide insight into how dynamical frictional processes shape the statistics of earthquakes.
This framework captures fault dynamics through a chain of $N$ blocks of mass $m$, each connected to its neighbors by springs of stiffness $k_c$ and to a slowly driven plate by springs of stiffness $k_p$. As the plate moves with velocity $v$, stress accumulates until blocks slip, producing earthquake-like events. Each block follows Newton’s equation of motion,
\begin{equation}\label{eqn:BK-Model_1}
m \ddot{u}_i = k_c \left( u_{i+1} - 2u_i + u_{i-1} \right) 
+ k_p \big( vt - u_i \big) - F_{\text{fric}}(\dot{u}_i),
\end{equation}
where the nonlinear friction force $F_{\text{fric}}(\dot{u_{i}})$ generates the characteristic stick–slip (SS) behavior, and is given by the expression:
\begin{align}\label{eqn:BK-Model_2}
    F_{\text{fric}}(\dot{u}_i) &=
    \begin{cases}
    F_s, & \text{if } \dot{u}_i = 0 \text{ and } |F_{\text{elastic}}| < F_s, \\[0.5em]
    \frac{1 - \sigma}{1 + \dfrac{2 \mu \, v}{1 - \sigma}}
, & \text{if sliding occurs},
    \end{cases}
    \end{align}
where $F_s$ is the static friction force, $\mu $\ is related to the friction weakening rate, and $\sigma$\ is related to the friction drop parameter, when it changes from static to kinetic. 
    
Using the stick–slip friction law with parameters $\delta$ (friction weakening rate) and $\sigma$ (friction drop), the model can reproduce Gutenberg–Richter–like magnitude statistics, exhibit chaotic dynamics despite being deterministic, and transition between periodic, chaotic, and scale-free regimes depending on parameters. Although more sophisticated rate-and-state (R\&S) friction laws exist, we use the stick–slip rule which is computationally simpler.

\section{Simulation Framework}\label{sec:sim}
For both the kinetic wealth exchange models: \textbf{B-Model} and \textbf{C-Model} described above, we simulate the models with $N=100$ agents and perform $1000$ independent realizations for each parameter configuration using Eqns.~\eqref{eqn:B-Model}, \eqref{eqn:B_with_l} and \eqref{eqn:C_with_l} respectively. 
Considering $w_i(t)$ be the wealth of agent $i$ at time $t$, the total wealth $W=\sum_{i=1}^{N} w_i(t)$ is always conserved.  
We initialize wealth as $w_i(0)=1$ for all $i$, and evolve the system for sufficiently long time ($t=10^6$) to ensure that a stationary distribution is reached. 
Inequality measures are computed from the steady-state wealth distributions averaged across realizations. 

In the \textbf{P-Model}, the agent incomes are drawn from a Pareto distribution defined on
\([1,\infty)\), with the minimum income set to unity and shape parameter
\(\alpha\), described using Eqn.~\eqref{eqn:Pareto}. Simulations are carried out for a population of \(10^6\) agents and
repeated over five independent realizations, with all reported quantities
obtained by averaging over these realizations to reduce finite-size
fluctuations.

For the \textbf{CS-Model}, the fractal representation of a fault surface is
constructed using the classical base-3 Cantor set to model the geometrical
overlap dynamics between two self-similar rough interfaces. For a given
generation $n$, the frequency of each overlap magnitude is evaluated using
Eqn.~\eqref{eqn:CS Model}, with $n$ ranging from 4 to 200. The inequality measures
$g$, $k$, and $p$ are computed from the corresponding Lorenz curves, where the cumulative normalized overlap magnitude (overlap magnitude at an instance / total number of overlaps over 3\(^n\) time steps) is plotted along the ordinate against the cumulative fraction of events (each instance of an overlap per time step, arranged in increasing order of overlap magnitude) along the abscissa. 
These measures provide a quantitative characterization of the degree of heterogeneity in the time series of overlap magnitudes.

For the \textbf{BK-Model}, we perform numerical simulations to investigate the emergence of collective slip dynamics for a given set of physical parameters.
The simulations are carried out under frictional and elastic conditions specified by the friction coefficient \(\mu\), the spring constants \(k_p\) and \(k_c\), and the number of blocks \(N\). 
At each time step, the instantaneous force acting on each block is computed using Eqns.~\eqref{eqn:BK-Model_1} and \eqref{eqn:BK-Model_2}. Using these forces, the elastic energy associated with the spring extensions is evaluated for all blocks according to the right-hand side of Eqn.~\eqref{eqn:BK-Model_1}. The corresponding Lorenz curve is then constructed by plotting the cumulative fraction of the total elastic energy released (obtained from these spring extensions) along the y-axis, against the cumulative fraction of blocks along the x-axis.
The degree of heterogeneity in the distribution of released energy is quantified using inequality indices, which serve as compact measures of the non-uniformity in the spring forces and provide a time-resolved characterization of the buildup and release of stress heterogeneity within the block ensemble.

\section{Results And Numerical Analysis}\label{sec:4}

For both the kinetic wealth exchange models, the P-model, and the earthquake-simulating models, we numerically evaluate the inequality indices the Gini index ($g$), the Pietra index ($p$) and the Kolkata index ($k$), along with their interrelations with different varying parameters and report the results in Tables~\ref{tab:1}--\ref{tab:5} in Appendices~\ref{app1} and~\ref{app2}. In Appendix ~\ref{app3} we plot the probability distribution function of wealth in the B-Model with varying range $R$, and the released elastic energy in the BK-Model at different times. 

\subsection{\textbf{B-Model with Exchange Range}}\label{sec:B-results}

\begin{figure}[t]
\centering    \includegraphics[width=0.33\linewidth]{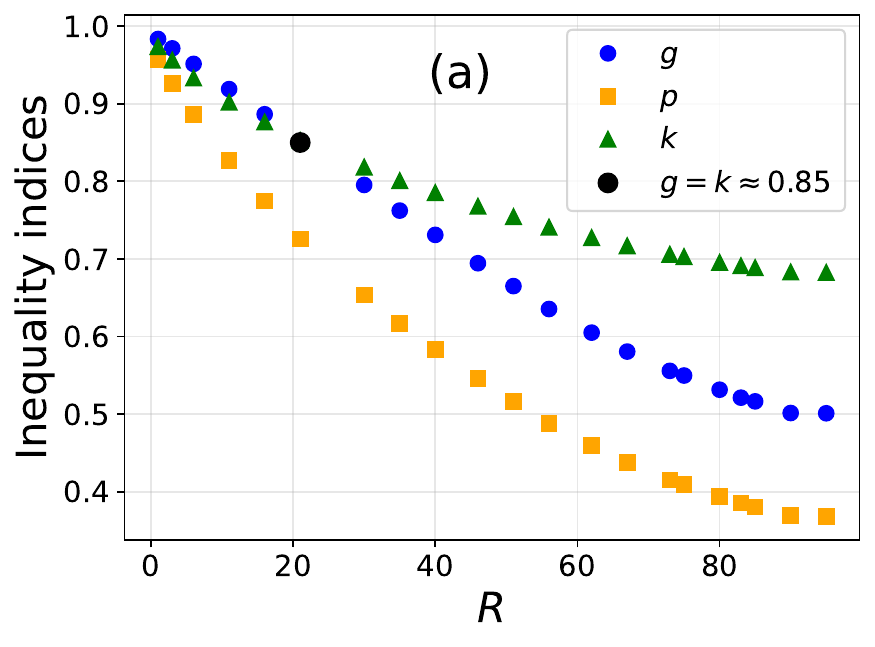}    \includegraphics[width=0.33\linewidth]{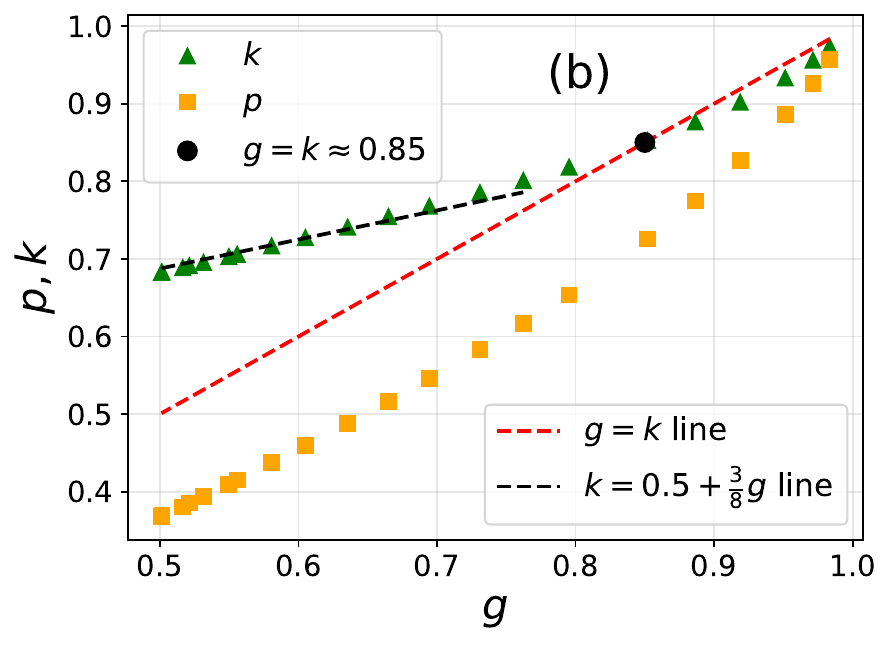}   \includegraphics[width=0.33\linewidth]{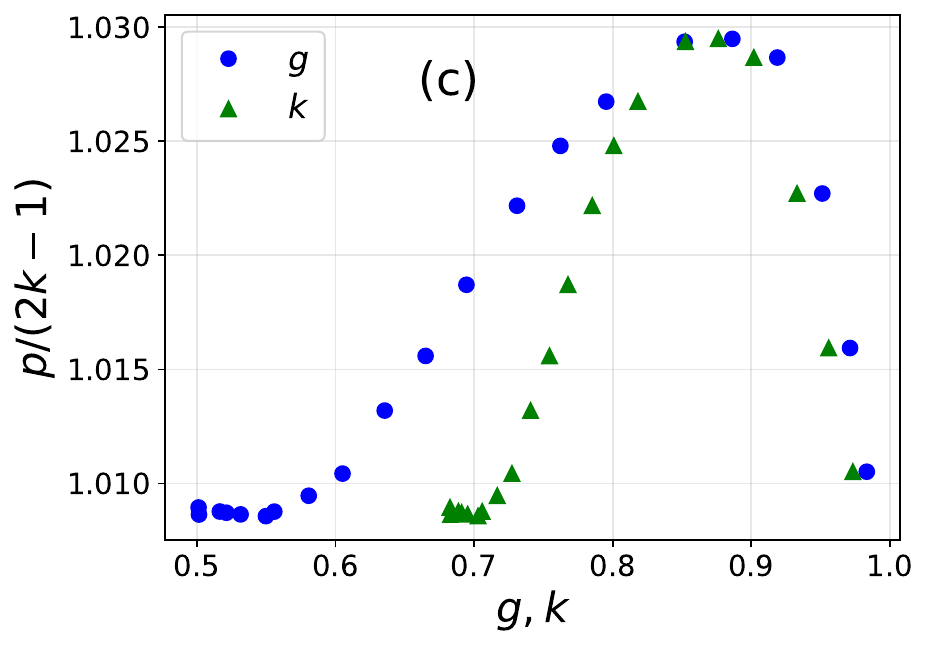}   \includegraphics[width=0.33\linewidth]{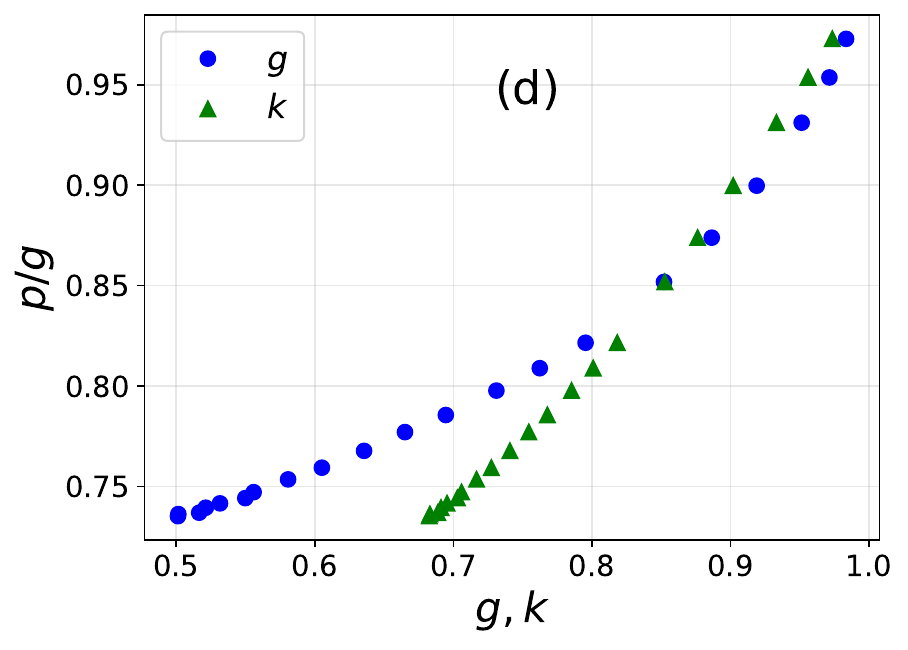}
\caption{\textbf{B-Model with Exchange Range:} Results of numerical analysis of the variations of inequality indices Gini ($g$), Pietra ($p$) and Kolkata ($k$) indices with range of exchanges ($R$) for B-Model (with total population, $N=100$ and ensemble averages over $10^3$ realizations).  (a) Variation of $g$, $p$ and $k$ with range $R$; (b) $g$ vs $k$-index curve showing initial straight line fitting with slope $3/8$ and the $g=k\approx 0.85$ point, also $g$ vs $p$-index curve; (c) Change of $p/(2k-1)$ with $g$ and $k$; (d) Change of $p/g$ with $g$ and $k$.}
    \label{fig:B_model}
\end{figure}

In this study, our primary objective is to examine three well-known measures of economic inequality and to investigate how they relate to one another within the framework of the B-Model as we vary the exchange range parameter $R$. Using Eqns.~\eqref{eqn:lor}--\eqref{eqn:k-index}, all three inequality indicators $g$, $p$, and $k$ are derived from the Lorenz curve ($L(x)$), which characterizes the cumulative distribution of wealth across the population following Eqn.~\eqref{eqn:B-Model}.

For each value of the exchange range parameter ($R$) used in the B-Model, we calculate all three indices and present the results in Table~\ref{tab:1} (see Appendix~\ref{app1}). Additionally, motivated from a recent work~\cite{asim_2025}, we examine the ratios $p/(2k-1)$ and $p/g$ to understand how these derived quantities evolve with the $g$ and $k$, offering further insights into the internal consistency and comparative behavior of these inequality measures.

Figure \ref{fig:B_model} provides a visual summary of how the inequality measures evolve as we vary the {\it range} parameter in the B-Model. In the upper-left panel, we can see the variation of $g$, $p$ and $k$ against $R$. In the upper-right panel, we illustrate the joint behaviors of the $g$-$k$ and $g$-$p$ pairs. As $R$ increases, $g$ and $k$ indices initially change in an almost perfectly linear manner with a slope $(3/8)$ as discussed in~\cite{bijin}. During this early phase, the values of $g$ and $k$ coincide at a point approximately $0.85$. Beyond this point, however, the relationship becomes distinctly nonlinear, indicating that the two measures respond differently to further increases in the range parameter. On the other hand as $R$ increases, $g$ and $p$ both increases together in similar manner and going to coincide at extreme inequality point $1$. 

The lower-left panel of Figure \ref{fig:B_model} displays the variation of the ratio $p/(2k-1)$ with respect to $g$ and $k$. 
The pattern shows an initial increase of this ratio up to a certain threshold value; subsequently, as $g$ and $k$ continues to grow, the ratio begins to decline.
Throughout the range, the ratio remains slightly above unity, reaching its maximum about a $3\%$ elevation near the point where $g$ and $k$ coincide (in this case, $g=k\sim 0.85$). 
This location has previously been identified as a precursor to criticality in several self-organized critical models~\cite{sand,manna,soumya}. 
The mild, non-monotonic rise toward this peak followed by a gradual decrease reflects a subtle shift in the balance between the $p$ and $k$ indices as the underlying wealth-exchange dynamics approach this near-critical region.
A similar trend is observed in the lower-right panel, which shows how the ratio $p/g$ depends on $g$ and $k$ indices. In this case, the ratio increases sharply with increasing the indices $g$ and $k$, suggesting that $p$ grows at a faster rate relative to $g$.
All these inequality indices are numerically computed from the steady-state wealth distributions for B-Model with different parameter (Exchange Range $R$), which are shown in Fig.~I in Appendix~\ref{app3}.
In the figure, we can clearly see that as $R$ increases the distribution goes to an exponential one but when $R=1$, we get a clear power-law with an exponential cut-off like distribution.

Now we can also show that varying the total population size does not shift the critical point, which remains located at approximately $g=k\sim 0.85$. However, the exchange range $R=R_c$ at which this critical point is reached depends on the population size. In the main analysis, we consider the total population $N=100$; additional results for $N=50$ and $N=200$ (presented in Fig. III(a) of Appendix~\ref{app4}) are analyzed.
For larger values of $N$, the dynamics become significantly slower, making it computationally difficult to reach the critical region. Therefore, we restrict the analysis to these three representative population sizes. As shown in Table~\ref{tab:extra}, the value of $R_c$ increases with increasing $N$, while the critical crossing point remains approximately unchanged at $g=k\sim 0.85$.
\begin{table}[H]
    \centering
    \small
    \begin{tabular}{|c|c|c|c|}
    \hline
    $N$&$R_c$&$g$&$k$\\ 
        \hline
    50&  10&  0.856	& 0.856 \\
        \hline
    100&  21&  0.852&  0.853 \\
         \hline
    200&  45&  0.850& 0.850\\
         \hline
  \end{tabular}
    \caption{\textbf{B-Model with Exchange Range:} Numerically estimated values, computed from the Eqn~\eqref{eqn:B-Model}, of the range $R=R_c$ at which $g=k$ for different populations $N$ for B-Model with Exchange Range (see Appendix~\ref{app4}).}
    \label{tab:extra}
\end{table}

\subsection{\textbf{B-Model with Savings}}

\begin{figure}[t]
\centering
\includegraphics[width=0.33\linewidth]{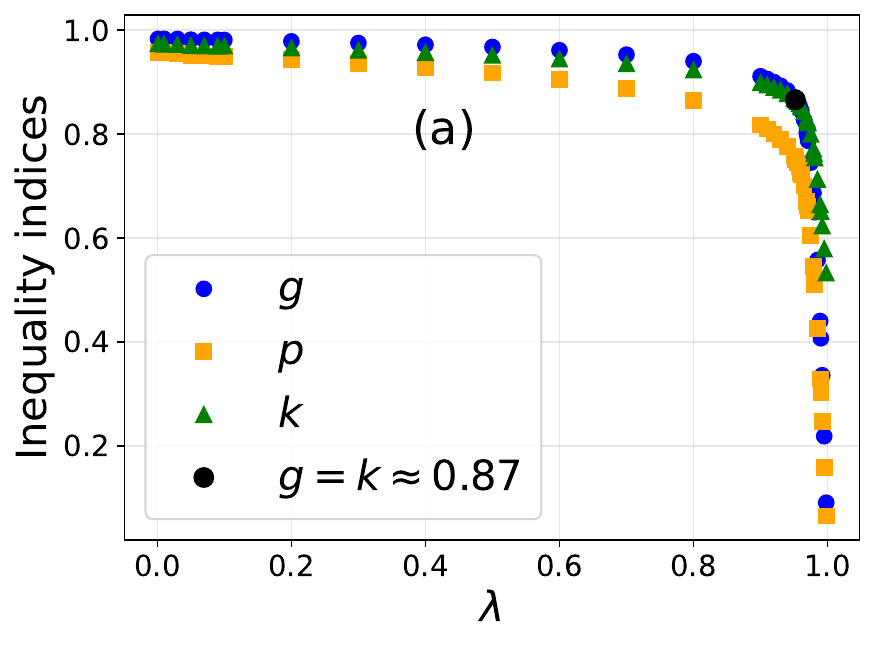}
\includegraphics[width=0.33\linewidth]{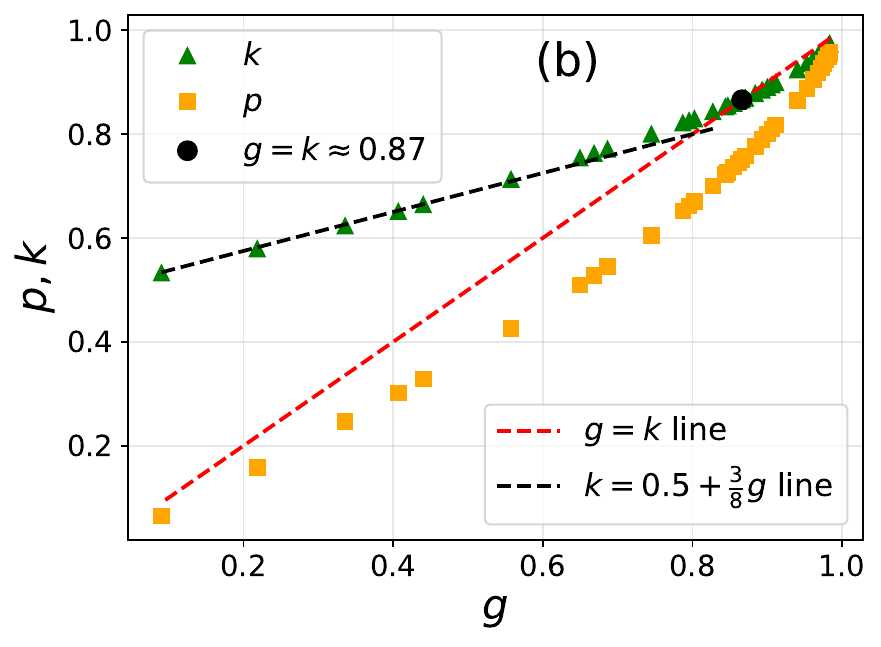}
\includegraphics[width=0.33\linewidth]{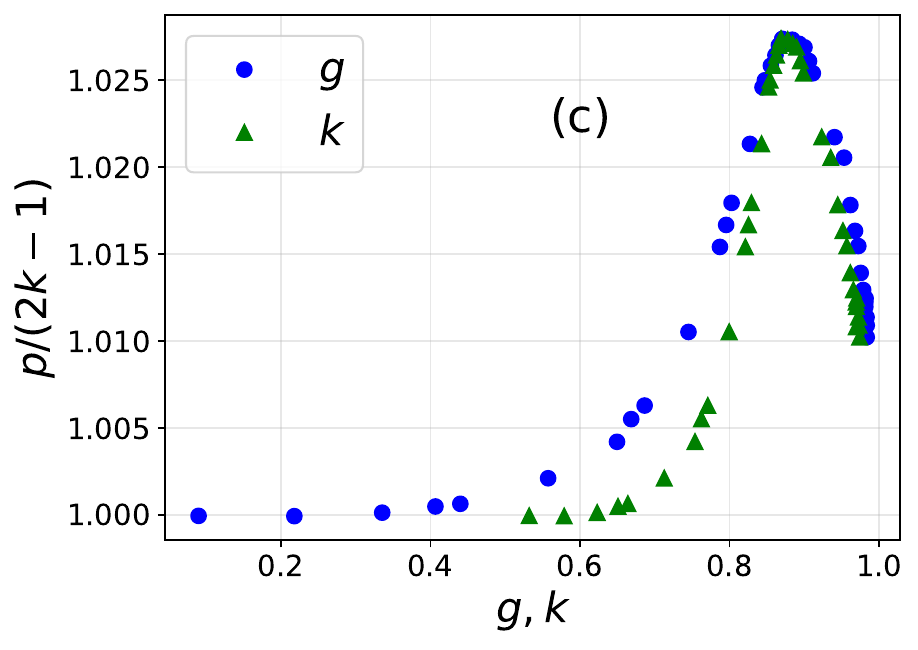}
\includegraphics[width=0.33\linewidth]{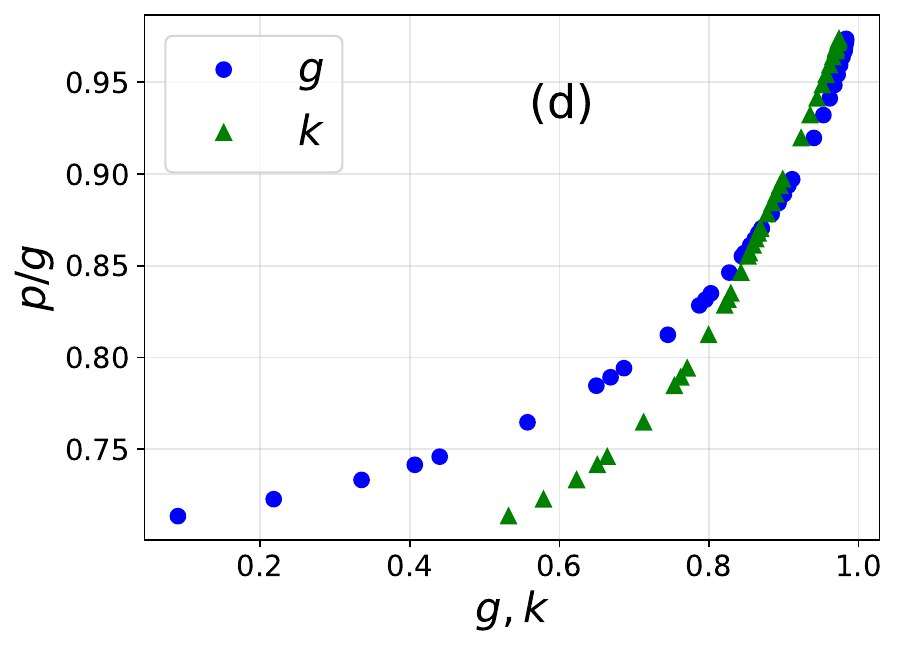}
\caption{\textbf{B-Model with Savings:} Results of numerical analysis of the variations of inequality indices Gini ($g$), Pietra ($p$) and Kolkata ($k$) indices with saving propensity ($\lambda$) for B-Model (with total population, $N=100$ and ensemble averages over $10^3$ realizations).  (a) Variation of $g$, $p$ and $k$ with savings $\lambda$; (b) $g$ vs $k$-index curve showing initial straight line fitting with slope $3/8$ and the $g=k\approx 0.87$ point, also $g$ vs $p$-index curve; (c) Change of $p/(2k-1)$ with $g$ and $k$; (d) Change of $p/g$ with $g$ and $k$.}
    \label{fig:B_with_l}
\end{figure}

In this case as well, we evaluate the same three inequality measures $g$, $p$ and $k$ to investigate how they behave and relate to one another when the model dynamics are controlled by the saving propensity $\lambda$ using Eqn.~\eqref{eqn:B_with_l}. 
Table~\ref{tab:2} in Appendix~\ref{app1} presents the computed values of $g$, $p$, and $k$ for a range of $\lambda$ in the B-Model (fixed the exchange range $R=1$) using Eqns.~\eqref{eqn:gini}, \eqref{eqn:pietra} and \eqref{eqn:k-index} respectively derived from Eqn.~\eqref{eqn:lor}. Alongside these primary indicators, we also examine the derived ratios $p/(2k-1)$ and $p/g$, which provide additional insight into how the internal structure of inequality evolves as agents save a larger fraction of their wealth. By observing how these quantities vary with $g$ and $k$, we can better understand the sensitivity of the model’s inequality characteristics to changes in saving propensity.

In the upper-left panel of Figure~\ref{fig:B_with_l}, we present the relationship between $g$, $p$ and $k$ as the saving propensity $\lambda$ varies. The upper-right panel of the figure depicts the variation of $k$ and $p$ with respect to $g$. Similar to the earlier case with the range parameter, $g$ and $k$ indices exhibit an almost linear relationship in the initial regime with slope $(3/8)$. They coincide at approximately $g\approx k \approx 0.87$, after which the curves begin to diverge, indicating a departure from the initial linear dependence as savings become more dominant in the wealth-exchange dynamics. Further we can see that $p$ is also increasing with $g$ starting both from $0$ and coincide at $1$.

The bottom-left panel of Figure~\ref{fig:B_with_l} illustrates how the ratio $p/(2k-1)$ evolves with $g$ and $k$. 
Consistent with our previous observation, the ratio remains slightly above unity across all values of $\lambda$, reaching a maximum of about a $3\%$ elevation near the point where $g$ and $k$ coincide (in this case $g=k\approx 0.87$). 
For small values of $g$ and $k$, the ratio increases gradually up to this peak. 
However, as these indices approach their coincident value near $0.87$, the ratio drops sharply. 
This behavior underscores the sensitivity of the interplay between the $p$ and $k$ indices as the system approaches this near-critical regime.
The bottom-right panel of the figure depicts the variation of the ratio $p/g$ as a function of $g$ and $k$. Here, a clear increasing trend emerges: as $g$ and $k$ increases, the ratio $p/g$ consistently increases, following the pattern illustrated in the plot. This monotonic increase indicates that $p$ grows more rapidly relative to $g$.

Similar to the total population $N=100$, we perform the same analysis for $N=50$ and $N=200$ (presented in Fig. III(b) of Appendix~\ref{app4}).
As we increase $N$ and also saving propensities $\lambda$, it will take long time to reach the critical point computationally. Here we can see in Table~\ref{tab:extra2}, the critical value of $\lambda=\lambda_c$ increases with increase in $N$, while the critical crossing point of $g$ and $k$ remains approximately at $0.86$.
\begin{table}[H]
    \centering
    \small
    \begin{tabular}{|c|c|c|c|}
    \hline
    $N$&$\lambda_c$&$g$&$k$\\ 
        \hline
    50&  0.848&	0.860&	0.860\\
        \hline
    100&  0.952& 0.866& 0.866\\
         \hline
    200&  0.970& 0.861& 0.861\\
         \hline
  \end{tabular}
    \caption{\textbf{B-Model with Savings:} Numerically estimated values, computed from the Eqn.~\eqref{eqn:B_with_l}, of $\lambda=\lambda_c$ at which $g=k$ for different populations $N$ for B-Model with Savings (see Appendix~\ref{app4}).}
    \label{tab:extra2}
\end{table}

\subsection{\textbf{C-Model with Savings}}\label{sec:C-results}

\begin{figure}[t]
\centering
\includegraphics[width=0.33\linewidth]{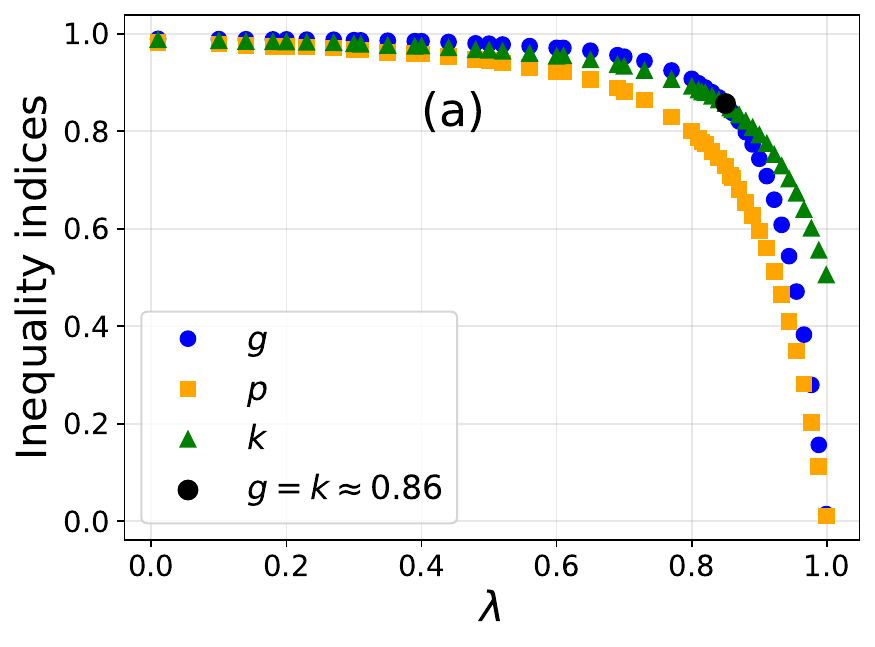}
\includegraphics[width=0.33\linewidth]{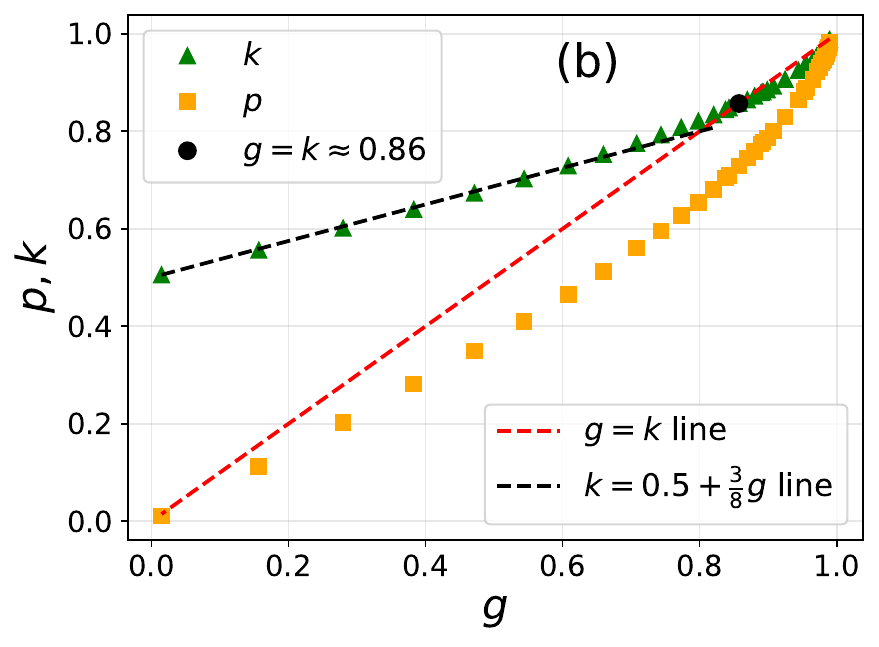}
\includegraphics[width=0.33\linewidth]{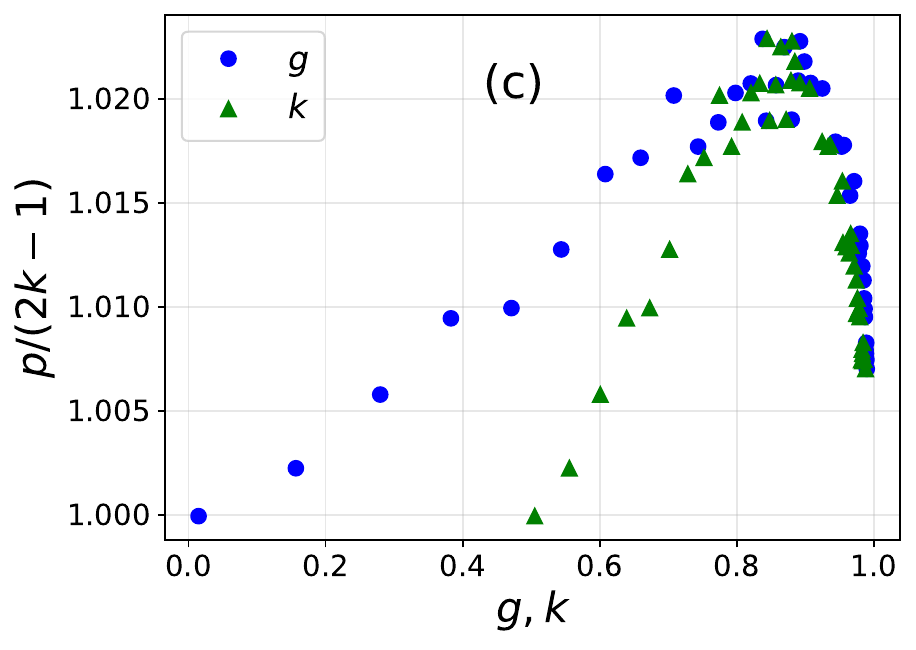}
\includegraphics[width=0.33\linewidth]{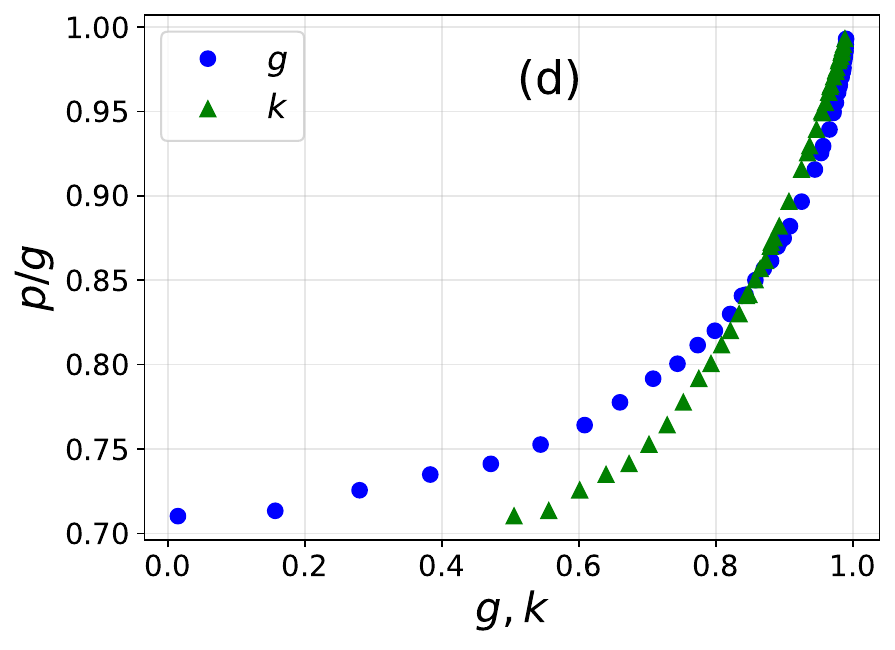}
\caption{\textbf{C-Model with Savings:} Results of numerical analysis of the variations of inequality indices Gini ($g$), Pietra ($p$) and Kolkata ($k$) indices with saving propensity ($\lambda$) for C-Model (with total population, $N=100$ and ensemble averages over $10^3$ realizations). (a) Variation of $g$, $p$ and $k$ with saving propensity $\lambda$; (b) $g$ vs $k$-index curve showing initial straight line fitting with slope $3/8$ and the $g=k\approx 0.86$ point, also $g$ vs $p$-index curve; (c) Change of $p/(2k-1)$ with $g$ and $k$; (d) Change of $p/g$ with $g$ and $k$.}
    \label{fig:YS_with_l}
\end{figure}

In this scenario as well, we evaluate the same set of three inequality indicators to understand how inequality evolves within the C-Model in presence of saving propensity, $\lambda$ following Eqn.~\eqref{eqn:C_with_l}. Table~\ref{tab:3} in Appendix~\ref{app1} reports the numerical values of these indices for various choices of $\lambda$ using Eqns.~\eqref{eqn:lor}--\eqref{eqn:k-index}. Along with the primary measures, we also examine the derived ratios $p/(2k-1)$ and $p/g$, which provide additional insight into the structural changes in inequality as the saving propensity is varied. By tracking how these ratios respond to changes in $g$ and $k$, we gain a clearer picture of the internal consistency and comparative behavior of the inequality measures in the C-Model.

The upper-left panel of Figure~\ref{fig:YS_with_l} illustrates how $g$, $p$ and $k$ co-varies with the saving propensity $\lambda$ in the C-Model or Yard-Sale Model framework. Further in the upper-right panel of this figure, we observe that at low and intermediate values of $g$, $k$ evolve in near synchrony, producing an almost linear trend. They intersect at a point close to $g = k \approx 0.86$. Beyond this intersection, however, the relation begins to bend, signaling a clear departure from linearity as higher saving propensities increasingly influence the wealth-exchange dynamics. Here we can also see the variation of $p$ with $g$, these two measures coincide only at $0$ and $1$.

The lower-left panel shows the behavior of the ratio $p/(2k-1)$ as a function of $g$ as well as $k$. 
The ratio remains slightly above unity, reaching a modest peak—about a $3\%$ increase near the point where $g$ and $k$ coincide. 
For small values of $g$ and $k$, the ratio increases gradually before attaining this peak. 
As the system approaches the coincident value (here $g=k\approx 0.86$), however, the ratio undergoes a pronounced decline, indicating a sharp reduction in the disparity between the $p$ and $k$ indices.
The lower-right panel displays the ratio $p/g$ across $g$ and $k$. Here, a clear monotonic increase is observed: as $g$ and $k$ increases, the value of $p/g$ steadily increases. This pattern indicates that $p$ becomes progressively larger relative to $g$ consistent with the trend captured in the plot.

In this model also we analyze the results for $N=50$ and $N=200$ (presented in Fig. III(c) of Appendix~\ref{app4}).
Again for larger populations and also for saving propensities the dynamics become significantly slower, making it computationally difficult to reach the critical region. As shown in Table~\ref{tab:extra1}, the critical value of $\lambda=\lambda_c$ decreases with increase in $N$, while the critical crossing point of $g$ and $k$ remains approximately unchanged at $0.85$.
\begin{table}[H]
    \centering
    \small
    \begin{tabular}{|c|c|c|c|}
    \hline
    $N$&$\lambda_c$&$g$&$k$\\ 
        \hline
    50&  0.895& 0.850&	0.850\\
        \hline
    100&  0.850& 0.857& 0.857\\
         \hline
    200&  0.795& 0.850&	0.850\\
         \hline
  \end{tabular}
    \caption{\textbf{C-Model with Savings:} Numerically estimated values, computed from the Eqn.~\eqref{eqn:C_with_l}, of $\lambda=\lambda_c$ at which $g=k$ for different populations $N$ for C-Model with Savings (see Appendix~\ref{app4}).}
    \label{tab:extra1}
\end{table}

\subsection{\textbf{P-Model}}\label{sec:P-Results}

\begin{figure}[t]
\centering
\includegraphics[width=0.33\linewidth]{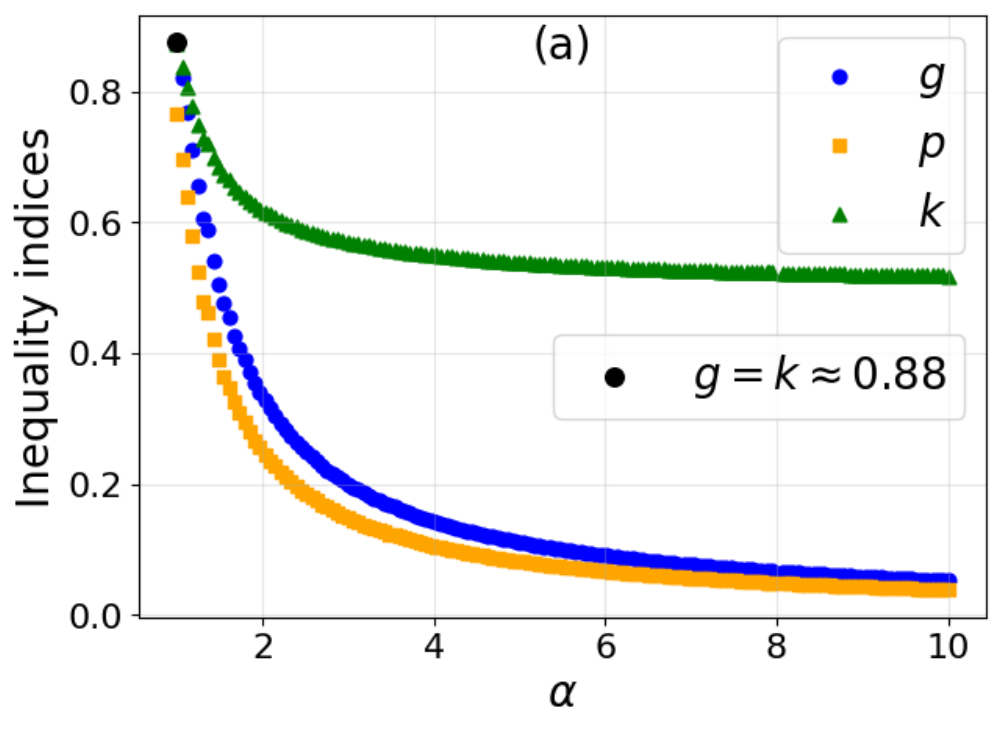}
\includegraphics[width=0.33\linewidth]{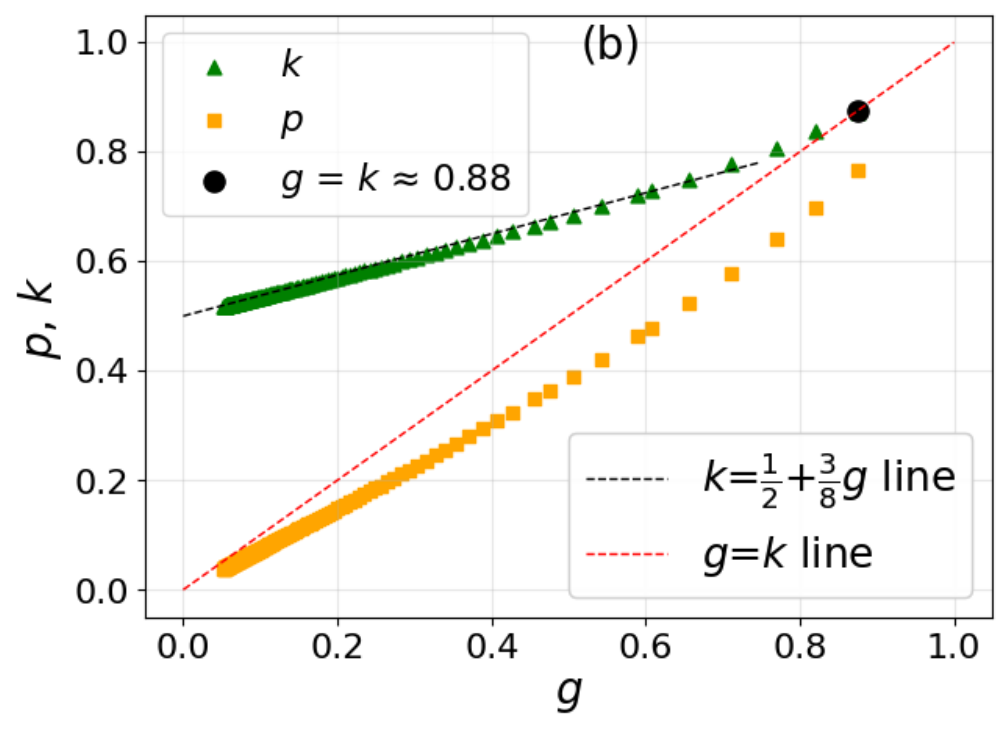}
\includegraphics[width=0.33\linewidth]{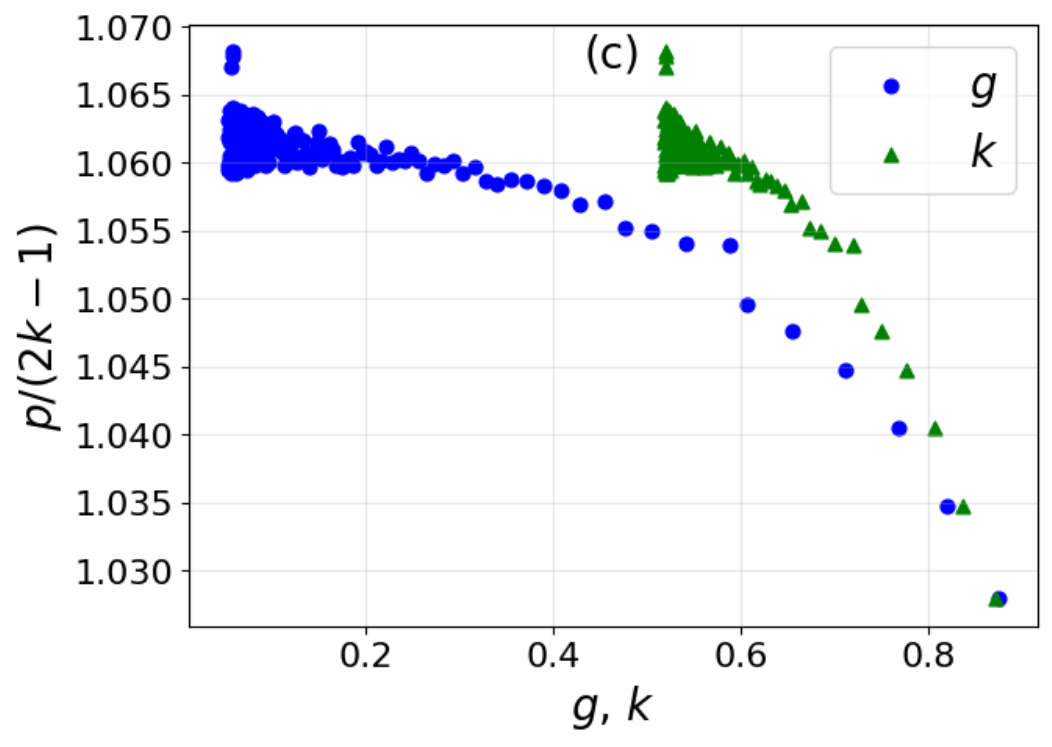}
\includegraphics[width=0.33\linewidth]{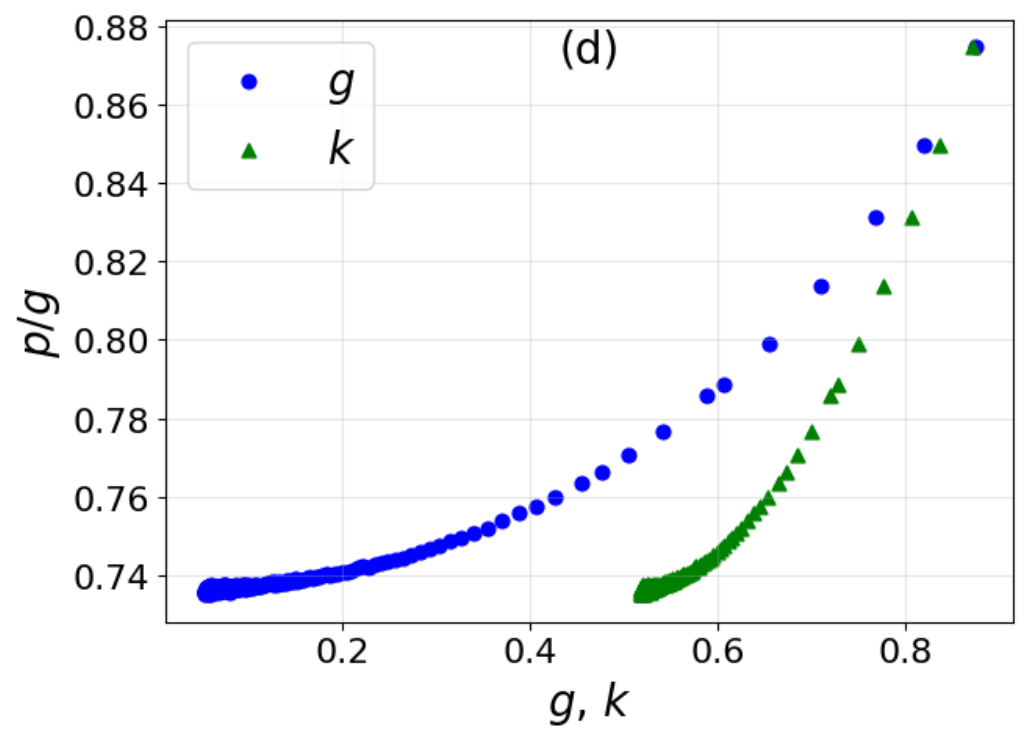}
\caption{\textbf{P-Model:} Results of numerical analysis of the variations of inequality indices Gini ($g$), Pietra ($p$) and Kolkata ($k$) indices with ($\alpha$) for P-Model for minimum income=$1$. (a) Variation of $g$, $p$ and $k$ with exponent $\alpha$; (b) $g$ vs $k$-index curve showing initial straight line fitting with slope $3/8$ and the $g=k\approx 0.88$ point, also $g$ vs $p$-index curve; (c) Change of $p/(2k-1)$ with $g$ and $k$; (d) Change of $p/g$ with $g$ and $k$.}
    \label{fig:pareto}
\end{figure}

We simulated income distributions governed by the Pareto law following Eqns.~\eqref{eqn:Pareto} and \eqref{eqn:lor_p} and evaluated the associated inequality measures- namely, the Gini coefficient $g$, the Pietra index $p$, and the Kolkata index $k$ using Eqns.~\eqref{eqn:lor}--\eqref{eqn:k-index}. 
Although analytic expressions exist for the inequality indices of the curve, we compute them numerically here to assess how well the simulation results agree with the analytic predictions. The minimum income is fixed at unity, and all three indices exhibit a systematic and monotonic dependence on the Pareto shape parameter $\alpha$. As $\alpha$ approaches its lower bound of $1^{+}$, the distribution enters the regime of maximal permissible inequality while still maintaining a finite mean. Correspondingly, the indices converge toward their characteristic limiting values.
A decrease in $\alpha$ enhances the heaviness of the power-law tail, implying an increasingly concentrated allocation of income among a diminishing fraction of individuals. This redistribution toward the upper tail is quantitatively captured by the pronounced rise in both $g$ and $k$ indices, each of which serves as a robust scalar measure of inequality. 
The computed values of the inequality indices with different values of $\alpha$ presents in Table~\ref{tab:4} shown in Appendix~\ref{app2}.

The upper-left panel of Figure~\ref{fig:pareto} shows how the inequality measures 
$g$, $p$, and $k$ evolve with the exponent $\alpha$. In the upper right panel we can see that $k$ and $g$ mostly follow a linear relation, till they both become equal at $\approx 0.88$, at $\alpha= 1.001$. We also see the $p$ is always lesser than $g$ for the values of $\alpha$ considered. 

The lower-left panel shows the behavior of the ratio $p/(2k-1)$ as a function of $g$ as well as $k$.
The ratio is always greater than 1, with its value decreasing from 
$\approx 1.03$ to $\approx 1.07$.
The lower-right panel displays the ratio $p/g$ across $g$ and $k$. A clear monotonic increase is observed.

\subsection{\textbf{CS-Model}}\label{sec:CS-Results}

\begin{figure}[t]
  \centering
\includegraphics[width=0.33\linewidth]{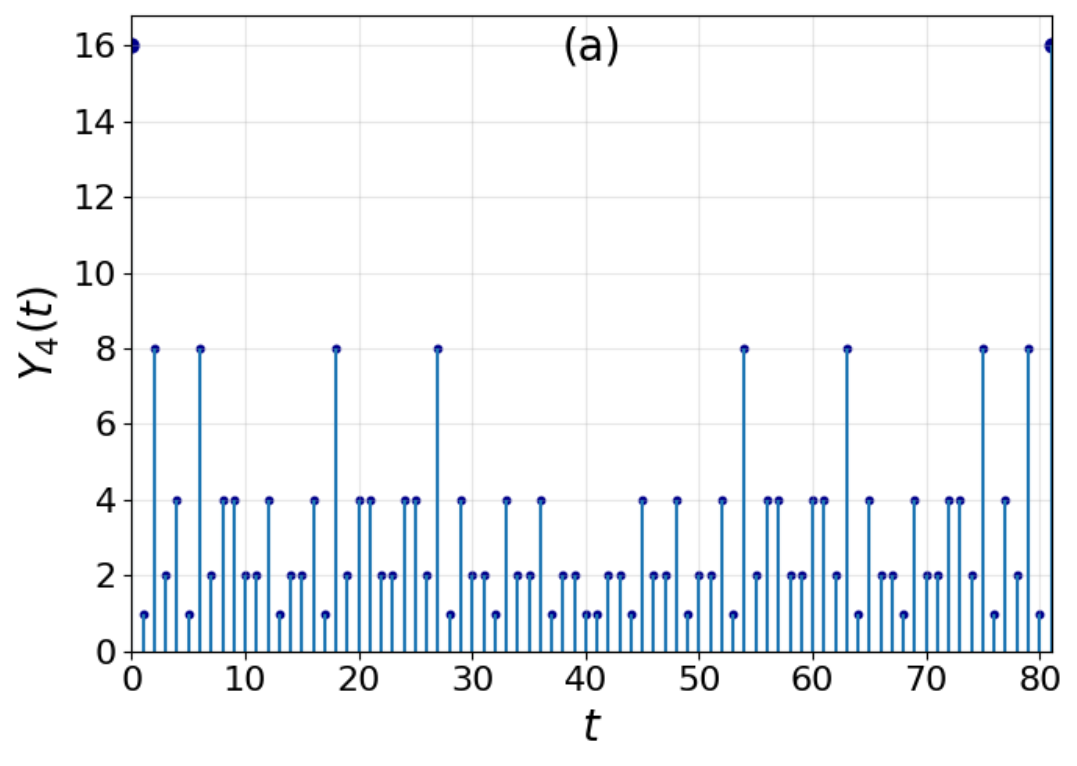}
  \includegraphics[width=0.33\linewidth]{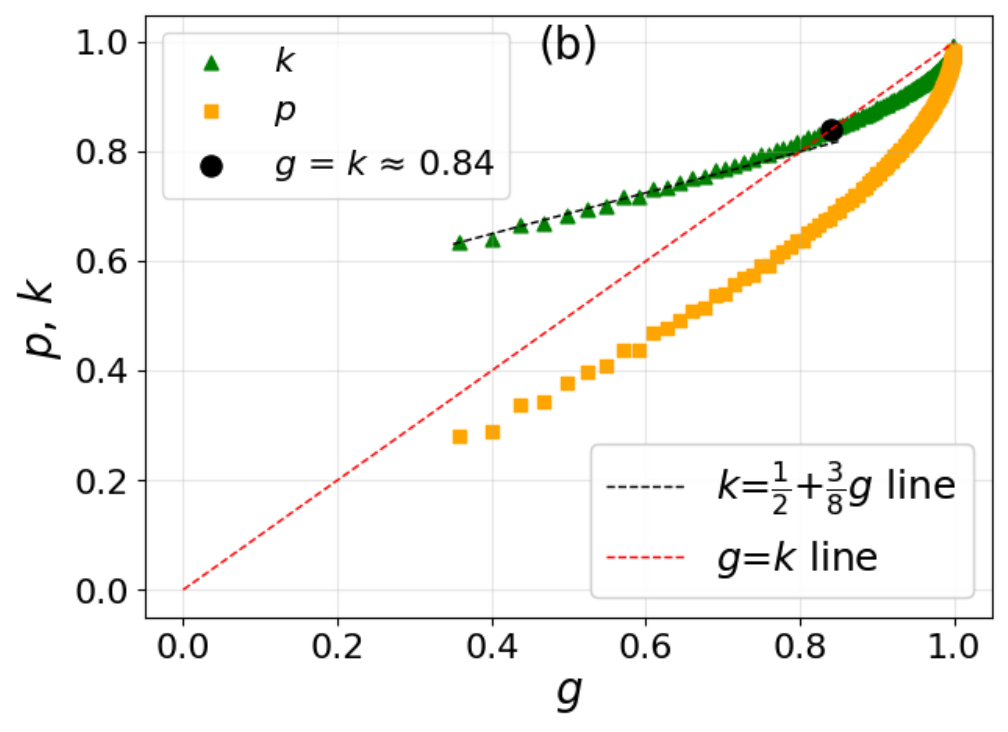}
  \includegraphics[width=0.33\linewidth]{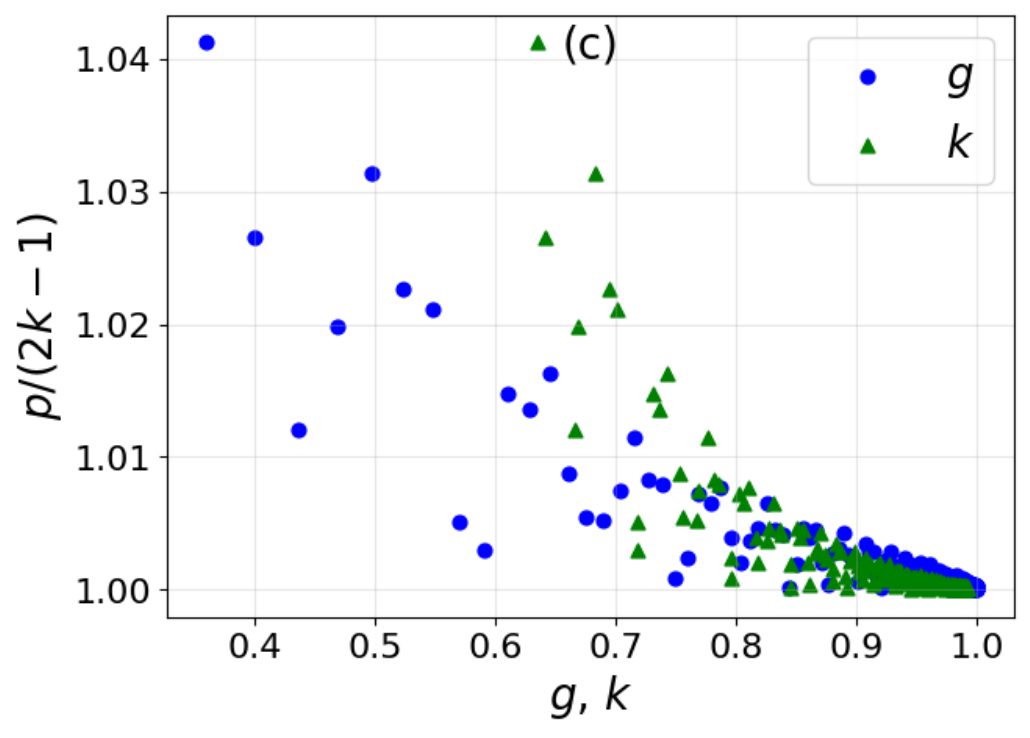}
  \includegraphics[width=0.33\linewidth]{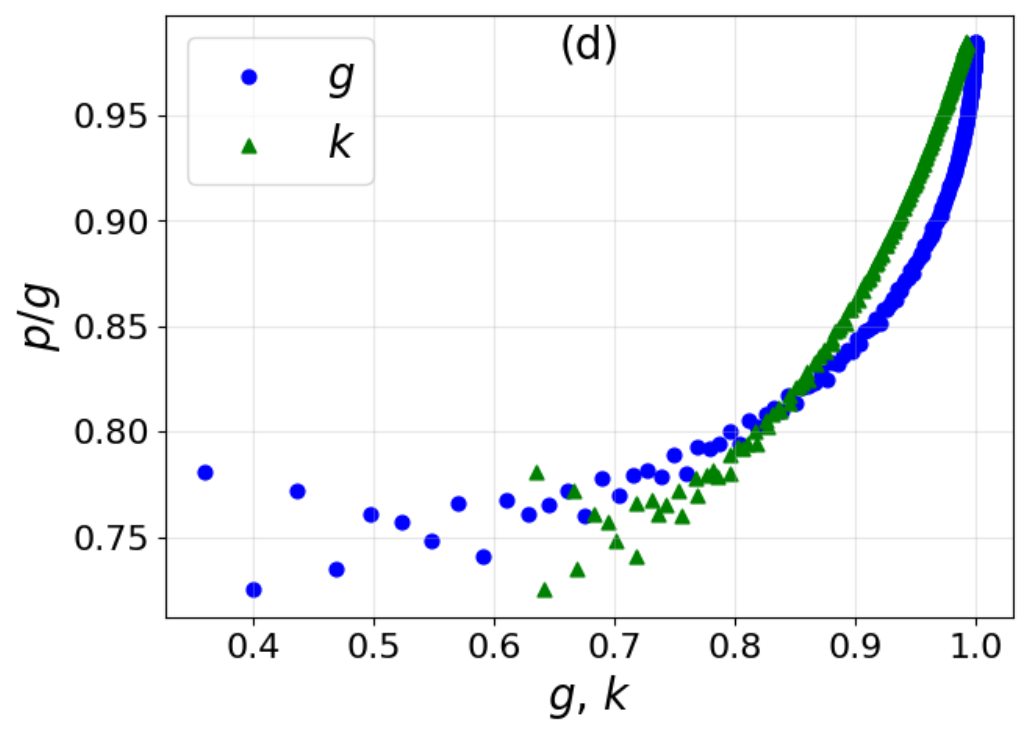}
      \caption{\textbf{CS-Model:} Results of numerical analysis of the variations of inequality indices Gini ($g$), Pietra ($p$) and Kolkata ($k$) indices with $n\in[4,200]$ for CS-Model. With increasing generation number, different values of $g$, $k$, and $p$ are obtained from the overlap time series. (a) Time ($t$) series of overlap magnitude $Y_4(t)$ of the 4th generation of the cantor set; (b) $g$ vs $k$-index curve showing initial straight line fitting with slope $3/8$ and the $g=k\approx 0.84$ point, also $g$ vs $p$-index curve; (c) Change of $p/(2k-1)$ with $g$ and $k$; (d) Change of $p/g$ with $g$ and $k$.}
    \label{fig:CS}
\end{figure}

The Kolkata--Gini relation obtained from the sequence of Cantor generations, initially exhibits small but systematic oscillations around an otherwise smooth and approximately monotonic curve. Here too, we calculate $g$, $k$ and $p$, along with the ratios $p/(2k-1)$ and $p/g$, to understand how inequality evolves with each fractal generation. The table \ref{tab:4} in Appendix~\ref{app2} shows values at 5-generation intervals from generation 5 onward, with generations 4 and 199 included separately.

The upper-left panel of Figure~\ref{fig:CS} shows how the inequality measures 
$g$, $p$, and $k$ evolve with the generation number in the CS-Model. 
The upper-right panel further reveals that, for low to intermediate values 
of $g$, the Kolkata index $k$ tracks $g$ closely, resulting in an almost 
linear movement. The two curves intersect near $g = k \approx 0.84$. 
Beyond this point, the relationship becomes non linear. The panel also displays how the Pietra index $p$ 
varies with $g$; these two measures coincide only at the maximal inequality value $p = g = 1$.

The lower-left panel shows the behavior of the ratio $p/(2k-1)$ as a function of $g$ as well as $k$. 
The ratio is always greater than 1, with its value decreasing from 
$\approx 10\%$ until it reaches 1 for the first time when 
$g = k \approx 0.84$.
The lower-right panel displays the ratio $p/g$ across $g$ and $k$. Here too, a clear monotonic increase is observed.

\subsection{\textbf{BK-Model}}\label{sec:BK-Results}

\begin{figure}[t]
  \centering
  \includegraphics[width=0.33\linewidth]{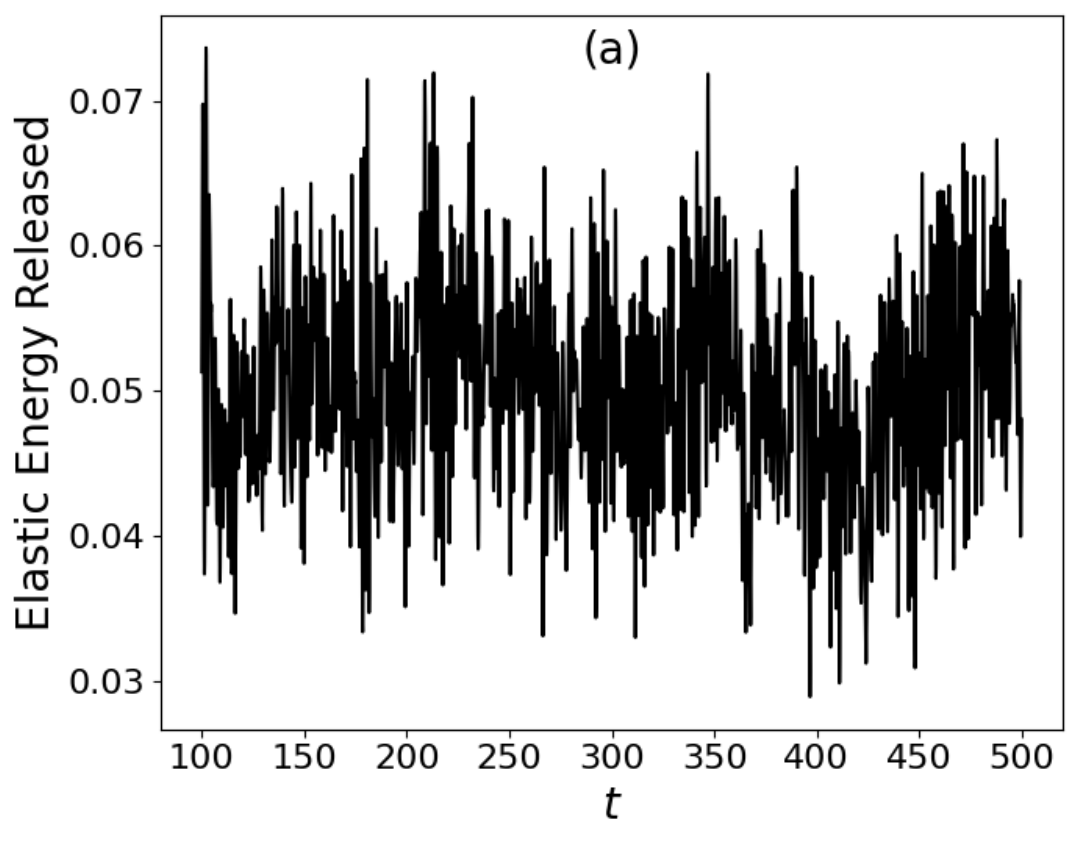}
  \includegraphics[width=0.33\linewidth]{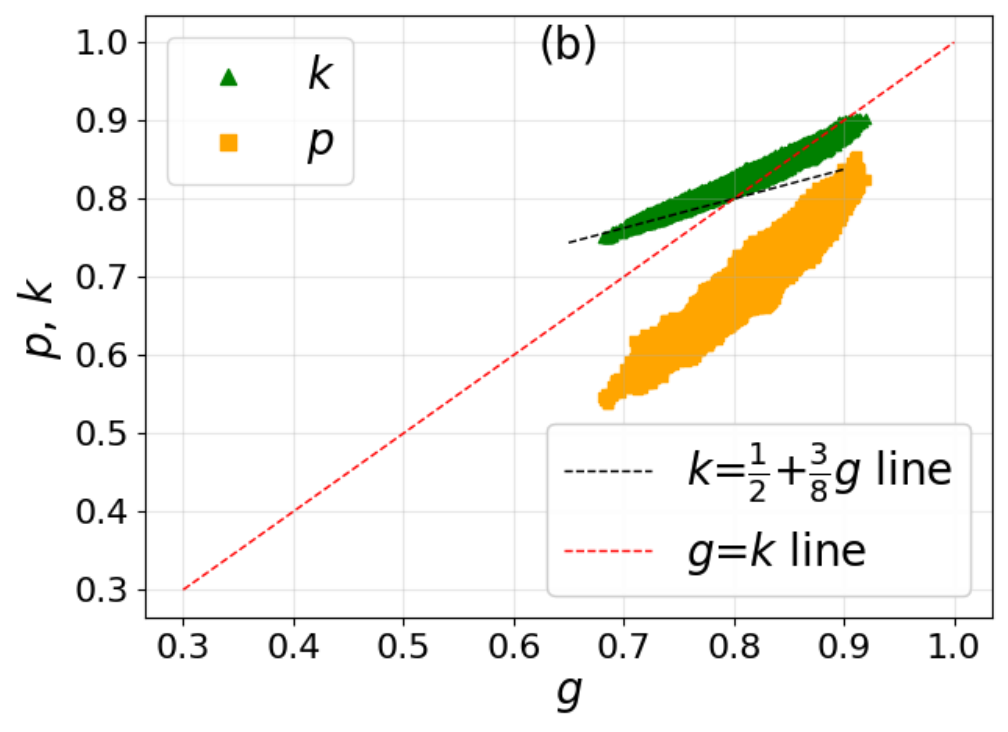}
  \includegraphics[width=0.33\linewidth=0.33]{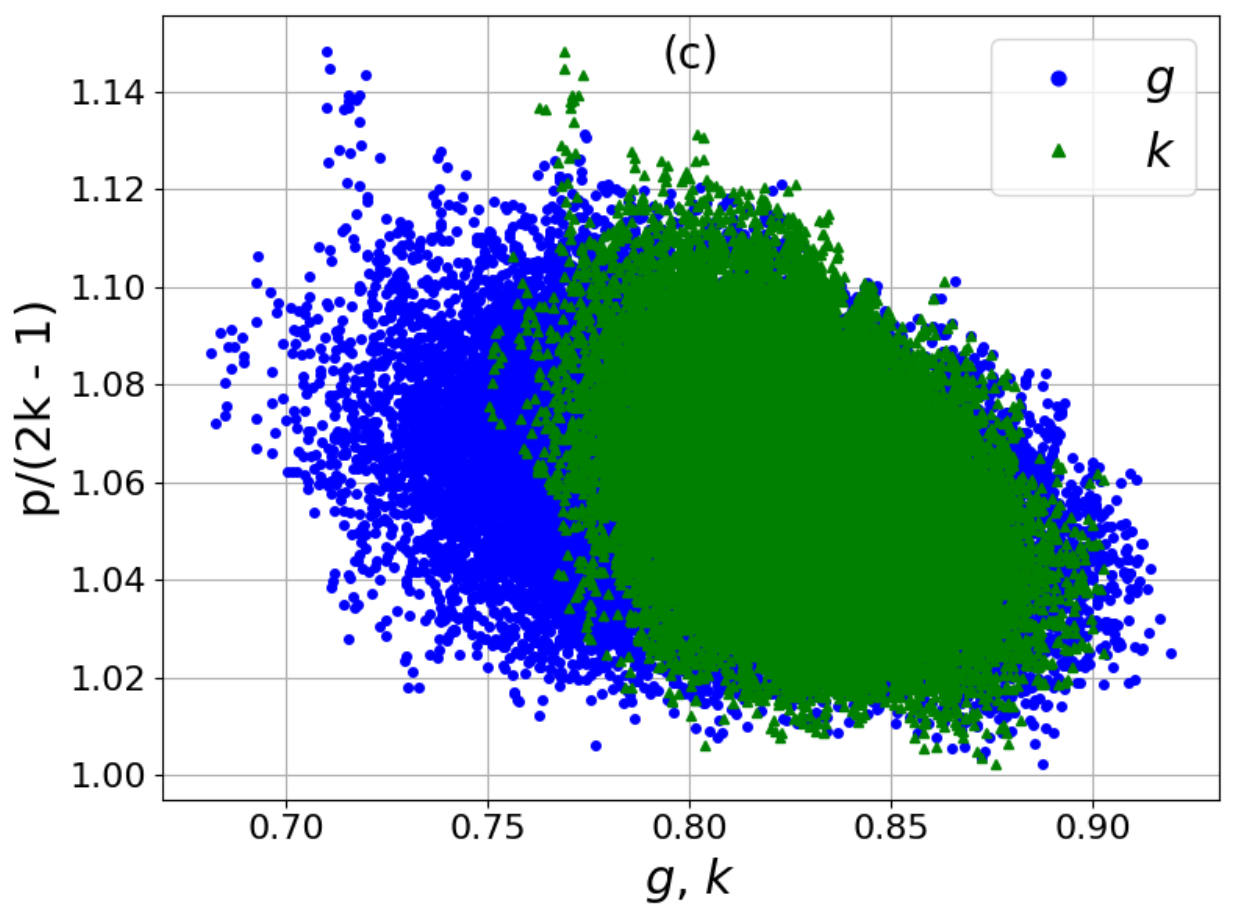}
  \includegraphics[width=0.33\linewidth]{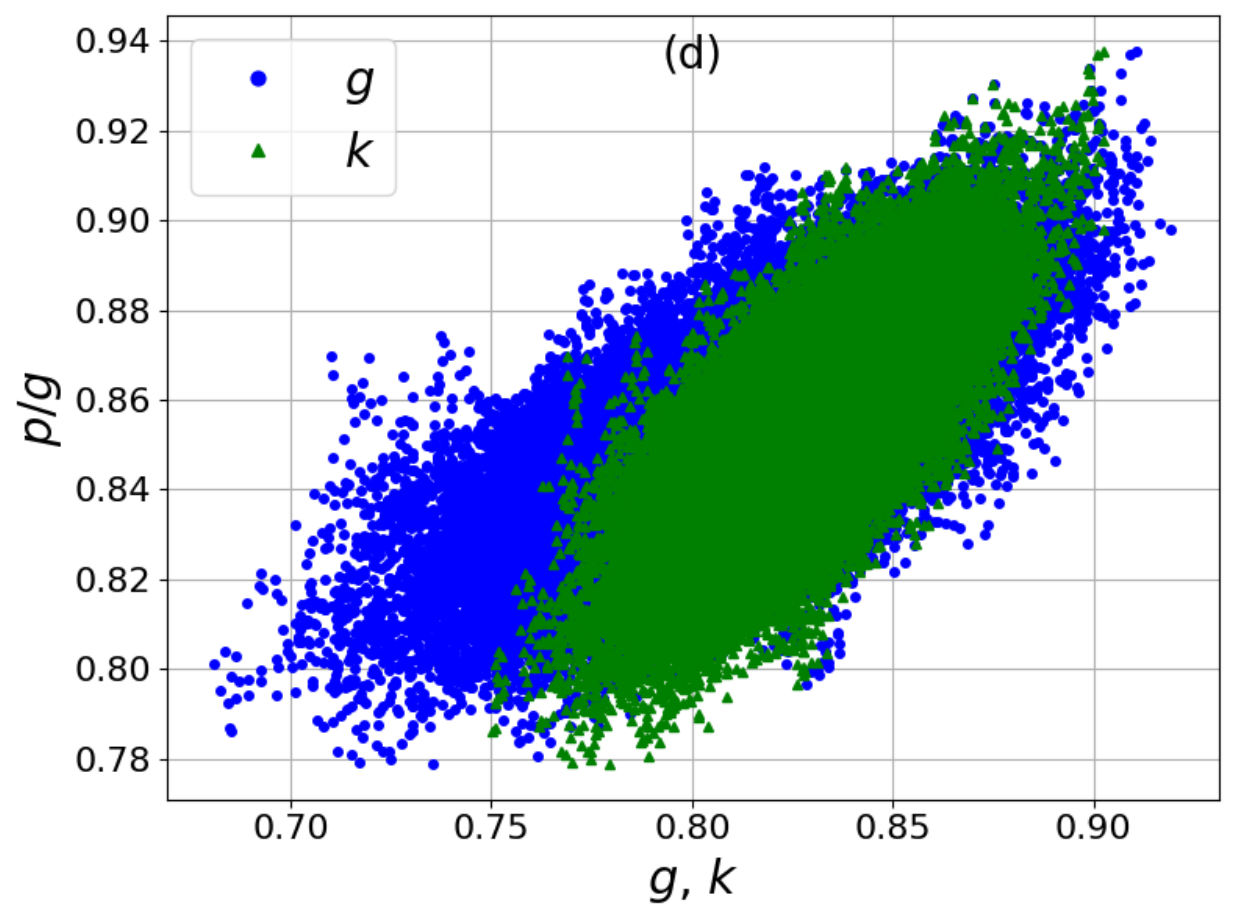}
\caption{\textbf{BK-model:} Results of numerical analysis of the variations of inequality indices Gini ($g$), Pietra ($p$) and Kolkata ($k$) indices with number of blocks ($N$) = $100$, $\mu = 1000$, $k_c/k_p= 1$ and $\sigma = 0.01$ for the BK-Model. (a) Elastic energy (potential energy stored in the springs) released with respect to time ($t$); (b) $g$ vs $k$-index curve showing initial straight line fitting with slope $3/8$ and also $g$ vs $p$-index curve; (c) $p$/$(2k-1)$ vs. $g$ and $k$ and; (d) $p$/$g$ vs. $g$ and $k$ plotted beyond the transient phase, after discarding the initial 100 time units, so the events are sampled from a stable distribution having both driving and dissipation. Here the equation of motion of the blocks and the frictional force acting on them are given by Eqns.~\eqref{eqn:BK-Model_1} and \eqref{eqn:BK-Model_2} respectively.} 
  \label{fig:BK}
\end{figure}

We analyze the temporal evolution of the elastic energy released and the
corresponding \(k\)–\(g\) and \(p\)–\(g\) relations. For the BK-Model, the \(k\)
versus \(g\) plot exhibits an approximately linear trend with slope
\(\simeq 0.375\), suggesting a nontrivial correspondence with the Lorenz-curve
geometry observed in the P-model, despite the fundamentally different underlying
dynamics. In the BK-Model, the inequality measures quantify the instantaneous distribution of spring forces across the block ensemble, providing a time-resolved characterization of stress heterogeneity during loading and slip
events. The released-energy time series in Figure \ref{fig:BK} represents the total energy released by the system, while $g$, $k$, and $p$ are calculated from the heterogeneity in spring-wise energy release at each time step. The time series of $g$, $k$, $p$ display an increase in their values immediately before the onset of a large slip event, as noted in \cite{unequalearthquakes_soumya}.

In the top right panel, we plot $k$ and $p$ with respect to $g$. In this regime of the model, $k$ and $g$ approximately follow a linear relationship with slope $3/8$ for values of $g$ upto 0.65. The points intersect the line of equality in multiple regions where $g$ ranges from $\approx 0.8$ to $\approx 0.9$. 

In the bottom left panel we show $p/(2k-1)$ which ranges from $\approx 1.00$ to $\approx 1.14$. In the bottom right panel, we plot $p/g$ where the ratio $p/g$ ranges from $\approx0.78$ and to $\approx0.94$. All the plots consist of points taken after the initial transient period (first 100 steps) has passed and a steady state is reached. In the Burridge-Knopoff model, the initial evolution is dominated by stress-loading
transients arising from the imposed initial conditions. During this phase, event
statistics and associated inequality measures
exhibit strong time dependence and are not representative of the intrinsic dynamics.

\section{Summary and concluding remarks}\label{sec:5}
In this study, we have performed a comparative numerical investigation of inequality measures (namely of Gini~\cite{gini}, Pietra~\cite{pietra} and Kolkata~\cite{k-index1,k-index3} indices) across a broad spectrum of model systems: ranging from two kinetic wealth exchange models (namely the Banerjee (B) Model~\cite{B_Model,B_Model1} and Chakraborti (C) model~\cite{C_Model} or, Yard-Sale model \cite{YS,YS1}), to synthetic Pareto (P) model~\cite{Pareto,Pareto1} and two earthquake-generating models (namely the Chakrabatrti- Stinchcombe (CS) model \cite{CS_Model,CS1,pbhattacharya} and the Burridge-Knopoff (BK) model~\cite{langer_1994,BK_Model}).
Using the Lorenz curves (see Fig.~\ref{fig:ineq}), generated numerically for each of these model systems for different parameter values (using Eqns.~\eqref{eqn:B-Model},\eqref{eqn:B_with_l} for B-Model, Eqn.~\eqref{eqn:C_with_l} for C-Model, Eqn.~\eqref{eqn:lor_p} for P-Model, Eqn.~\eqref{eqn:CS Model} for CS-Model and Eqns.~\eqref{eqn:BK-Model_1},\eqref{eqn:BK-Model_2} for the BK-Model), we estimate the different inequality indices, namely the Gini index ($g$), Pietra index ($p$), and Kolkata index ($k$).
It may be mentioned here, although the dynamics in the wealth distribution models (B and C), and in the earthquake models (CS and BK) are different, the inequality measures in ``wealth" possessed by different agents and in (effective) elastic energies associated with the different bursts or avalanches in the respective models in their steady states, show some remarkable universal feature. 
The results of our numerical analysis for the Gini index $g$ (given by the normalized area between the equality line and the Lorenz curve in Fig.~\ref{fig:ineq}, ranging from $0$ to $1$) and Kolkata index $k$ (given by the fixed point of the complementary Lorenz curve in Fig.~\ref{fig:ineq}, ranging from $1/2$ to $1$) in particular show that for all these models (B, C, CS and BK), both $g$ and $k$ become equal to about $0.86$ at their respective critical points (for B and C models) or self-organized critical points (for the CS and BK models). Note that $k = 0.80$ would correspond to Pareto's $80-20$ law, and $k = 0.86$ here means $86\%$ wealth (or energy) is possessed by (or released through) $14\%$ agents (or avalanches) at the critical or self-organised critical points for the respective models. Also, like the universality in the values of critical exponents near the critical points, the values of the inequality measures $g$ and $k$ show universality for the critical points of the B and C model (though the critical points can vary with the model parameters; see discussions in sections~\ref{sec:4}.B and \ref{sec:4}.C) and for the self-organized critical points of the CS and BK models.
The numerical results are all given in Tables~\ref{tab:1}-\ref{tab:5} (see Appendices~\ref{app1} and \ref{app2}) and Figs.~\ref{fig:B_model}-\ref{fig:BK}.
It may be noted that while the inequalities in the steady state wealth distributions in these models arise from the respective rules of kinetic exchanges among agents or traders, the unequal distribution  of  the two-fractal overlap measure (equivalent to the released elastic energy avalanches) over time in the CS model model follows  from the constant velocity slip dynamics of one fractal over the other and the steady state unequal elastic energy release avalanches in the BK model results from the relative motion of the linear springs and nonlinear friction forces. 
We note, however,  while switching from the econophysical wealth exchange models (B and C models) which have got dominantly exponential (to Gamma-like) distributions and the two earthquake simulating models (CS and BK models) which have got (Gutenberg-Richter like) dominantly power law distributions, the crossing (or Self-Organized Criticality-like) point for the inequality indices ($g$ and $k$) studied here compared well with that for the perfect power-law Pareto (P) distribution.

This study reveals a strikingly consistent crossing point between the inequality measures $g$ and $k$ across a diverse set of systems, including kinetic wealth exchange models, a synthetic Pareto model, and earthquake-generating models, under varying control parameters. In the B-Model, the crossing occurs around $g=k\approx0.85$ (see Fig.~\ref{fig:B_model}(b)) for different exchange ranges and shifts slightly to $g=k\approx0.87$ (see Fig.~\ref{fig:B_with_l}(b)) with varying savings. Similarly, the C-Model exhibits a crossing near $g=k\approx 0.86$ (see Fig.~\ref{fig:YS_with_l}(b)), while the Pareto model yields a value close to $g=k\approx0.88$ (see Fig.~\ref{fig:pareto}(b)) upon changing the shape parameter. For the earthquake models, the crossing point is approximately $0.84$ in the CS-Model and about 
$0.85$ in the BK-Model (see Figs.~\ref{fig:CS}(b) and \ref{fig:BK}(b)).
Taken together, these results for the inequality measures of the distributions (of wealth and avalanche) in the  socio-economic and geophysical models considered here,  point to a robust tendency for the measures $g$ and $k$ to converge toward a near-universal  value around $0.86\pm0.02$, aligning with earlier observations in self-organized critical systems~\cite{manna,soumya}, suggesting a deeper underlying connection between inequality growth dynamics and criticality. Next we have studied the values of $p/(2k-1)$ (across the wealth exchange models, the two-fractal model and the Pareto distributions) and found them to remain a little above unity, as was predicted theoretically~\cite{asim_2025}.
The predicted values (assuming a simple polynomial form for the Lorenz function~\cite{asim_2025}) of $p/g = 3/4$  seems to deviate quite a bit (in the range $0.69$ to about $1.0$; see Tables~\ref{tab:1}-\ref{tab:5} in Appendices~\ref{app1}-\ref{app2}), while the  existence of the relation~\cite{asim_2025} $k = 1/2 + (3/8)g$ seems to be valid for all the models considered here in the region of lower $g$ values (see Figs.~\ref{fig:B_model}(d) - \ref{fig:BK}(d)).

The broad range of models considered here is an essential part of the present study. By comparing wealth-exchange models, a synthetic Pareto distribution, and earthquake-generating models within the same Lorenz-curve framework, we show that the inequality indices $g$, $p$, and $k$ are not restricted to a single application domain. Rather, they provide a common statistical language for describing concentration phenomena in systems as different as wealth dynamics and avalanche dynamics. This is particularly significant because these measures are relevant both for characterizing wealth inequality and for identifying the buildup toward large events in earthquake-like systems. The near-universal crossing behavior of $g$ and $k$ therefore gains additional importance as a possible signature of generic organization in disparate non-equilibrium systems.
These observations of quantitatively similar behaviors for these inequality indices, studied here for different socio-economic and geophysical models, may provide some useful and coherent comparative framework for studying the universal statistical features of disparate dynamical systems.

\appendix

\section{Tables for B-Model and C-Model}\label{app1}

\begin{table}[H]
    \centering
    \small
    \begin{tabular}{|c|c|c|c|c|c|}
    \hline
    $R$&$g$&$p$&$k$&$p/(2k-1)$&$p/g$\\ 
        \hline
         1&  0.983&  0.957&  0.973 &1.011
&0.973 
\\
        \hline
         3&  0.971&  0.926&  0.956 &1.016
&0.954 
\\
         \hline
         6&  0.951&  0.886&  0.933& 1.023
&0.931 
\\
         \hline
         11&  0.919&  0.827&  0.902 & 1.028
&0.900 
\\
       \hline
      16&  0.886&  0.775&  0.876 &1.029 
&0.874 
\\
         \hline
        21&  0.852&  0.726&  0.853 & 1.030
&0.852
\\
         \hline
         30&  0.796&  0.653&  0.818 &1.027 
& 0.822 
\\
         \hline
         35&  0.762&  0.617&  0.801 &  1.025
&0.809 
\\
        \hline
         40&  0.731&  0.583&  0.785 & 1.022
& 0.798 
\\
        \hline
        46&  0.695&  0.546&  0.768 &1.019
&0.786 
\\ 
\hline
        51&  0.665&  0.517&  0.754 &1.016
&0.778 \\ 
        \hline
        \end{tabular}
         \begin{tabular}{|c|c|c|c|c|c|}
    \hline
    $R$&$g$&$p$&$k$&$p/(2k-1)$&$p/g$\\ 
        \hline
        56&0.636&0.488&0.741& 1.013
&0.768
\\ 
         \hline
    62&0.605&0.460&0.727& 1.010 
&0.759 
\\
        \hline
        67&0.581&0.438&0.717&1.009 
&0.754 
\\ 
        \hline
      73&0.556&0.415&0.706&1.008
&0.747  
\\ 
        \hline
      75&0.550&0.409&0.703&1.008  
&0.744 
\\ 
        \hline
      80&0.532&0.394&0.695& 1.009
&0.742 
\\ 
       \hline
        83&0.521&0.385&0.691&1.009
&0.739
\\
       \hline
        85&0.517&0.381&0.689&1.009 
&0.737 
\\
        \hline
        90&0.502&0.369&0.683&1.009
& 0.736 
\\
       \hline
        95&0.501&0.368&0.683&1.009
&0.735 
\\ 
    \hline
    100&0.501&0.368&0.683&1.010&0.737\\ 
       \hline
  \end{tabular}
    \caption{\textbf{B-Model with Exchange Range:} Numerically estimated values of Gini ($g$), Pietra ($p$) and Kolkata ($k$) indices for B-Model with different range $R$ computed from the Eqn~\eqref{eqn:B-Model}. Also showing the computed ratios $p/(2k-1)$ (motivated by \cite{asim_2025}) and $p/g$ values for B-model with different range ($R$).}
    \label{tab:1}
\end{table}
\begin{table}[H]
    \centering
    \small
    \begin{tabular}{|c|c|c|c|c|c|}
    \hline
    $\lambda$&$g$&$p$&$k$&$p/(2k-1)$&$p/g$\\ 
    \hline
     0.001&0.983 &0.957 &0.974  & 1.010&0.974\\
    \hline
     0.01& 0.983&0.957 &0.973 &1.011&0.973\\
     \hline
     0.03&0.983 &0.955 &0.972 &1.011&0.972\\
    \hline
     0.05& 0.982 & 0.951&0.970 &1.012 &0.970\\
    \hline
     0.07& 0.982&0.950 &0.970  &1.012 &0.968\\
     \hline
    0.09&0.981 & 0.949&0.969 &1.012& 0.968\\
     \hline
     0.10&0.981 &0.949 & 0.969&1.011&0.967\\
    \hline
    0.20& 0.979 & 0.943&0.966 &1.013&0.964\\
    \hline
   0.30& 0.975& 0.936&0.961 & 1.014&0.959\\ 
    \hline
   0.40&0.972&0.928&0.957&1.015&0.954\\ 
     \hline
    0.50&0.968&0.918&0.952&1.016&0.948\\
    \hline
    0.60&0.962&0.905&0.945&1.018&0.941\\ 
    \hline
    0.70&0953 &0.888&0.935&1.021&0.932\\ 
    \hline
    0.80&0.940&0.865 &0.923&1.022&0.920\\ 
    \hline
    0.90&0.911&0.818&0.899&1.025&0.897\\ 
   \hline
    0.91&0.906&0.810&0.895&1.026 &0.894\\
    \hline
    0.92&0.900&0.800&0.890&1.027&0.889\\
    \hline
    0.93&0.893&0.790&0.884&1.027 &0.884\\
    \hline
    0.94&0.884&0.776&0.878& 1.027&0.878\\
    \hline
    0.95& 0.871&0.758 &0.869&1.027 &0.870\\
    \hline
    \end{tabular}
    \begin{tabular}{|c|c|c|c|c|c|}
    \hline
    $\lambda$&$g$&$p$&$k$&$p/(2k-1)$&$p/g$\\ 
    \hline
    0.951&0.870&0.758&0.869&1.027 &0.870\\
    \hline
    0.952&0.866&0.751&0.866& 1.027&0.868\\
    \hline
    0.955&0.861&0.745&0.863&1.026 &0.864\\
    \hline
   0.958&0.855&0.737&0.859& 1.026&0.861\\
    \hline
    0.96&0.847&0.726&0.854&1.025 &0.857\\
    \hline
 0.961&0.844&0.722&0.852&1.025 &0.855\\
    \hline
   0.965&0.827&0.700&0.843&1.021&0.846\\
   \hline
    0.969&0.803&0.670&0.829&1.018&0.835\\
    \hline
    0.97&0.795&0.661&0.825&1.017 &0.832\\
    \hline
    0.971&0.787&0.652 &0.821&1.015&0.828\\
    \hline
    0.975& 0.745 & 0.605 &0.800&1.011 &0.812\\
    \hline
    0.979&0.686&0.545&0.771& 1.006&0.794\\
   \hline
    0.98&0.668 &0.528&0.762& 1.006&0.789\\
    \hline
    0.981& 0.650 &0.510&0.754& 1.004&0.785\\
    \hline
    0.985& 0.557&0.426 &0.713 &1.002&0.765\\
    \hline
    0.989&0.440&0.328  &0.664	&1.001 &0.746\\
    \hline
    0.99&0.407 &0.302 &0.651&1.001&0.742\\
    \hline
    0.992&0.336 &0.246 &0.623	& 1.000&0.733\\
    \hline
    0.995&0.218 &0.158 &0.579	&0.999 &0.723\\
    \hline
    0.998& 0.090 &0.064  &0.532 &0.999 &0.713\\
   \hline
  \end{tabular}
    \caption{\textbf{B-Model with Savings:} Numerically estimated values of Gini ($g$), Pietra ($p$) and Kolkata ($k$) indices for B-model with different saving propensities $\lambda$ computed from the Eqn.~\eqref{eqn:B_with_l}. Also showing the computed ratios $p/(2k-1)$ (motivated by~\cite{asim_2025}) and $p/g$ values for B-model with different saving propensities $\lambda$.}
    \label{tab:2}
\end{table}
\begin{table}[H]
    \centering
    \small
    \begin{tabular}{|c|c|c|c|c|c|}
    \hline
    $\lambda$&$g$&$p$&$k$&$p/(2k-1)$&$p/g$\\ 
        \hline
         0.01&  0.990&  0.983&  0.988 &1.007&0.993\\
        \hline
         0.10&  0.989&  0.979&  0.986 &1.008&0.990\\
         \hline
         0.14&  0.989&  0.977&  0.984 & 1.008&0.988\\
         \hline
         0.18&  0.989&  0.975&  0.984 & 1.008&0.986\\
        \hline
         0.20&  0.989&  0.974&  0.983 & 1.007&0.985\\
         \hline
         0.23&  0.988&  0.973&  0.983& 1.008&0.985\\
         \hline
         0.27&  0.988&  0.971&  0.982& 1.007&0.983 \\
         \hline
         0.30&  0.987&  0.968&  0.980& 1.009&0.981\\
        \hline
        0.31&  0.987&  0.967&  0.979& 1.010&0.980\\
        \hline
       0.35&  0.986&  0.962&  0.976& 1.010&0.976\\ 
        \hline
        0.39&0.985&	0.959&	0.974& 1.011&0.974\\ 
         \hline
        0.40&	0.984&	0.959&	0.975& 1.009&0.974\\
        \hline
        0.44& 0.983&0.954&	0.971& 1.011&0.970\\ 
        \hline
        0.48& 0.981&0.947&	0.967& 1.012&0.966\\ 
        \hline
        0.50& 0.980& 0.945&0.966& 1.013&0.965\\ 
        \hline
        0.52&	0.978&	0.940&	0.964&1.013&0.961\\ 
       \hline
        0.56&	0.975&	0.931&	0.960 & 1.013&0.955\\
        \hline
        0.60&	0.971&	0.923&	0.954& 1.016&0.950\\
        \hline
        0.61&	0.971&	0.922&	0.955& 1.013&0.949\\
        \hline
        0.65&	0.965&	0.907&	0.947& 1.015&0.939\\
        \hline
        0.69& 0.956 &	0.889 &	0.937 & 1.018&0.929\\
        \hline
        0.70& 0.953 &	0.882 &	0.933 & 1.018&0.925\\
        \hline
        0.73& 0.944 &	0.864 &	0.925 & 1.018&0.916\\
        \hline
        \end{tabular}
        \begin{tabular}{|c|c|c|c|c|c|}
        \hline
    $\lambda$&$g$&$p$&$k$&$p/(2k-1)$ & $p/g$\\ 
        \hline
        0.77& 0.925 &	0.829 &	0.906 & 1.020&0.897\\
        \hline 
        0.80& 0.908 &	0.801 &	0.892 & 1.020&0.882\\
        \hline
        0.81& 0.899 &	0.786 &	0.885 & 1.021&0.875\\
        \hline
        0.815& 0.892 &	0.778 &	0.881 & 1.022&0.872\\
        \hline
        0.82& 0.890 &	0.774 &	0.879 & 1.021&0.870\\
        \hline
        0.83& 0.880 &	0.758 &	0.872 & 1.019&0.862\\
        \hline
        0.84& 0.870 &	0.745 &	0.864 & 1.022&0.857\\
        \hline
        0.85& 0.857 &	0.729 &	0.857 & 1.020&0.850\\
        \hline
       0.857& 0.843 &	0.709 &	0.848 & 1.019&0.841\\
        \hline
       0.86& 0.837 &0.704 &	0.844 & 1.023&0.840\\
        \hline
       0.87& 0.821 &	0.681 &	0.834 & 1.021&0.830\\
        \hline
       0.88& 0.798 &0.655 &	0.821 & 1.020&0.820\\
        \hline
        0.89& 0.773 &	0.627 &	0.808 & 1.019&0.812\\
        \hline
        0.90& 0.744 &	0.595 &	0.792 & 1.018&0.801\\
        \hline
        0.911& 0.708 &	0.561 &	0.775 & 1.020&0.792\\
        \hline
        0.922& 0.660 & 0.513 &	0.752 & 1.017&0.778\\
        \hline
        0.933& 0.608 & 0.465 &	0.729 & 1.016&0.764\\
        \hline
        0.944& 0.544 & 0.409 &	0.702 & 1.013&0.753\\
        \hline
        0.955& 0.471 & 0.349 &	0.673 & 1.010&0.741\\
        \hline
        0.966& 0.383 & 0.281 &	0.639 & 1.009&0.735\\
        \hline
        0.977& 0.280 & 0.203 &	0.601 & 1.005&0.726\\
        \hline
        0.988& 0.157 & 0.112 &	0.556 & 1.002&0.713\\
        \hline
        0.999& 0.015 & 0.010 &	0.505 & 1 &0.710\\
       \hline
  \end{tabular}
    \caption{\textbf{C-Model with Savings:} Numerically estimated values of Gini ($g$), Pietra ($p$) and Kolkata ($k$) indices for C-model with different saving propensities $\lambda$ computed from the Eqn.~\eqref{eqn:C_with_l}. Also showing the computed ratios $p/(2k-1)$ (motivated by \cite{asim_2025}) and $p/g$ values for C-model with different saving propensities $\lambda$.}
    \label{tab:3}
\end{table}

\section{Tables for P-Model and CS-Model}\label{app2}
\begin{table}[H]
    \centering
    \small
    \begin{tabular}{|c|c|c|c|c|c|}
    \hline
    $\alpha$ & $g$ & $p$ & $k$ & $p/(2k-1)$ & $p/g$ \\ \hline 
    1.001&0.875&0.765&0.872&1.028&0.875\\ \cline{1-6} 1.243&0.655&0.524&0.750&1.047&0.799\\ \cline{1-6}1.545&0.476&0.365&0.673&1.055&0.766\\  \cline{1-6}  1.847&0.371&0.280&0.632&1.059&0.754\\  \cline{1-6}  2.149&0.304&0.227&0.607&1.059&0.747\\  \cline{1-6}  2.209&0.293&0.219&0.603&1.060&0.747\\  \cline{1-6}  2.451&0.256&0.590&0.190&1.060&0.743\\  \cline{1-6}  2.752&0.222&0.165&0.578&1.061&0.742\\  \cline{1-6}  3.054&0.196&0.145&0.568&1.061&0.740\\  \cline{1-6}  3.356&0.175&0.129&0.561&1.060&0.740\\   \cline{1-6}3.658&0.158&0.117&0.555&1.061&0.739\\ \cline{1-6}  3.960&0.144&0.107&0.550&1.061&0.738\\    \cline{1-6}    4.262&0.133&0.098&0.546&1.062&0.738\\  \cline{1-6}  4.564&0.123&0.091&0.543&1.062&0.738\\  \cline{1-6}  4.866&0.115&0.084&0.540&1.062&0.737\\  \cline{1-6}  
    5.168&0.107&0.079&0.537&1.062& 0.738\\    
    \hline
    \end{tabular}
    \begin{tabular}{|c|c|c|c|c|c|}
    \hline
    $\alpha$ & $g$ & $p$ & $k$ & $p/(2k-1)$ & $p/g$ \\ \hline  5.470&0.101&0.074&0.535&1.063&0.738\\  \cline{1-6}  5.772&0.095&0.070&0.533&1.061&0.738\\  \cline{1-6}  6.074&0.090&0.066&0.531&1.063&0.737\\  \cline{1-6} 6.376&0.085&0.063&0.529&1.063&0.737\\   \cline{1-6} 6.678&0.081&0.060&0.528&1.062&0.737\\  \cline{1-6} 6.980&0.077&0.057&0.527&1.062&0.737\\  \cline{1-6}  7.282&0.074&0.054&0.526&1.060&0.736\\  \cline{1-6}  7.584&0.071&0.052&0.524&1.061&0.736\\   \cline{1-6} 7.886&0.068&0.045&0.523&1.060&0.736\\  \cline{1-6}  8.188&0.065&0.048&0.523&1.060&0.736\\   \cline{1-6} 8.490&0.063&0.046&0.522&1.062&0.737\\  \cline{1-6}8.792&0.060&0.044&0.521&1.062&0.736\\  \cline{1-6}  9.094&0.058&0.043&0.520&1.062&0.737\\   \cline{1-6} 9.396&0.056&0.041&0.519&1.067&0.737\\   \cline{1-6} 9.698&0.054&0.040&0.519&1.064&0.737\\  \cline{1-6} 10.000&0.053&0.039&0.518&1.063&0.736\\
    \hline
    \end{tabular}%
        \caption{\textbf{P-Model:} Numerically estimated values of Gini ($g$), Pietra ($p$) and Kolkata ($k$) indices for P-model with different exponent $\alpha$, calculated using Eqn.~\eqref{eqn:lor_p}, along with the computed ratios $p/(2k-1)$ (motivated by~\cite{asim_2025}) and $p/g$ values for P-model with different exponent $\alpha$.}
    \label{tab:4}
\end{table}
\begin{table}[H]
    \centering
    \begin{tabular}{|c|c|c|c|c|c|}
    \hline
    $n$ & $g$ & $p$ & $k$ & $p/(2k-1)$ & $p/g$ \\ \hline  
    4&0.359&0.280 & 0.634 & 1.041&0.781\\
    \cline{1-6}
    5&0.400&0.290&0.641&1.026&   0.725\\   
    \cline{1-6}
    10&0.548&0.409&0.700&1.021&   0.748\\
    \cline{1-6}
    15&0.645&0.493&0.743&1.016&   0.765\\
    \cline{1-6}
    20&0.715&0.558&0.776&1.011&   0.779\\
    \cline{1-6}
    25&0.769&0.610&0.803&1.007&    0.793\\   
    \cline{1-6}
    30&0.811&0.653&0.825&1.004&   0.805\\   
    \cline{1-6}
    35&0.844&0.690&0.845&1.000&   0.817\\   
    \cline{1-6}
    40&0.871&0.721&0.860&1.002&    0.828\\   
    \cline{1-6}
    45&0.893&0.749&0.873&1.003&   0.839\\   
    \cline{1-6}
    50&0.911&0.773&0.885&1.003&    0.849\\   
    \cline{1-6}
    55&0.925&0.794&0.896&1.002&    0.858\\   
    \cline{1-6}
    60&0.937&0.812&0.906&1.001&    0.867\\   
    \cline{1-6}
    65&0.947&0.829&0.914&1.001&    0.875\\   
    \cline{1-6}
    70&0.956&0.845&0.922&1.001&    0.884\\   
    \cline{1-6}
    75&0.963&0.860&0.929&1.001&    0.893\\   
    \cline{1-6}
    80&0.968&0.873&0.936&1.001&    0.901\\   
    \cline{1-6}
    85&0.973&0.884&0.942&1.001&   0.908\\   
    \cline{1-6}
    90&0.977&0.895&0.947&1.001&    0.915\\   
    \cline{1-6}
    95&0.981&0.904&0.952&1.001&    0.921\\   
    \cline{1-6}
    100&0.984&0.912&0.956&1.000&   0.927\\ 
    \hline
    \end{tabular} 
    \begin{tabular}{|c|c|c|c|c|c|}
    \hline
    $n$ & $g$ & $p$ & $k$ & $p/(2k-1)$ & $p/g$ \\ \hline  
    105&0.986&0.920&0.960&1.000&   0.933\\
    \cline{1-6}
    110&0.988&0.926&0.963&1.000&   0.938\\   
    \cline{1-6}
    115&0.990&0.933&0.966&1.001&   0.942\\   
    \cline{1-6}
    120&0.992&0.938&0.969&1.000&   0.946\\   
    \cline{1-6}
    125&0.993&0.943&0.971&1.000&   0.950\\   
    \cline{1-6}
    130&0.994&0.948&0.974&1.000&   0.953\\   
    \cline{1-6}
    135&0.995&0.952&0.976&1.000&   0.957\\   
    \cline{1-6}
    140&0.996&0.956&0.978&1.000&   0.960\\   
    \cline{1-6}
    145&0.996&0.960&0.980&1.000&   0.963\\   
    \cline{1-6}
    150&0.997&0.963&0.981&1.000&   0.966\\   
    \cline{1-6}
    155&0.997&0.966&0.983&1.000&   0.969\\   
    \cline{1-6}
    160&0.998&0.969&0.984&1.000&   0.971\\   
    \cline{1-6}
    165&0.998&0.972&0.9860&1.000&   0.973\\   
    \cline{1-6}
    170&0.998&0.974&0.987&1.000&   0.976\\   
    \cline{1-6}
    175&0.998&0.976&0.988&1.000&   0.977\\   
    \cline{1-6}
    180&0.999&0.978&0.989&1.000&   0.979\\   
    \cline{1-6}
    185&0.999&0.980&0.990&1.000&   0.981\\   
    \cline{1-6}
    190&0.999&0.981&0.991&1.000&   0.982\\   
    \cline{1-6}
    195&0.999&0.982&0.991&1.000&   0.983\\ 
    \cline{1-6}
    199&0.999&0.984&0.992&1.000&0.984\\
    \cline{1-6}   
    200&0.999&0.984&0.992&1.000&   0.985\\ 
    \hline
   \end{tabular}
    \caption{\textbf{CS-Model:} Numerically estimated values of Gini ($g$), Pietra ($p$) and Kolkata ($k$) indices for CS-model with different generations ($n$), where the overlap magnitude $Y_n(t)$ is calculated using Eqn.~\eqref{eqn:CS Model} along with the computed ratios $p/(2k-1)$ and $p/g$ (motivated by~\cite{asim_2025}).}
    \label{tab:5}
\end{table}

\section{Probability distribution of wealth in B-Model with different Exchange Range $\mathbf{R}$ and elastic energy in the BK model at different times}\label{app3}
\begin{figure}[H]
    \centering
    \includegraphics[width=0.48\linewidth]{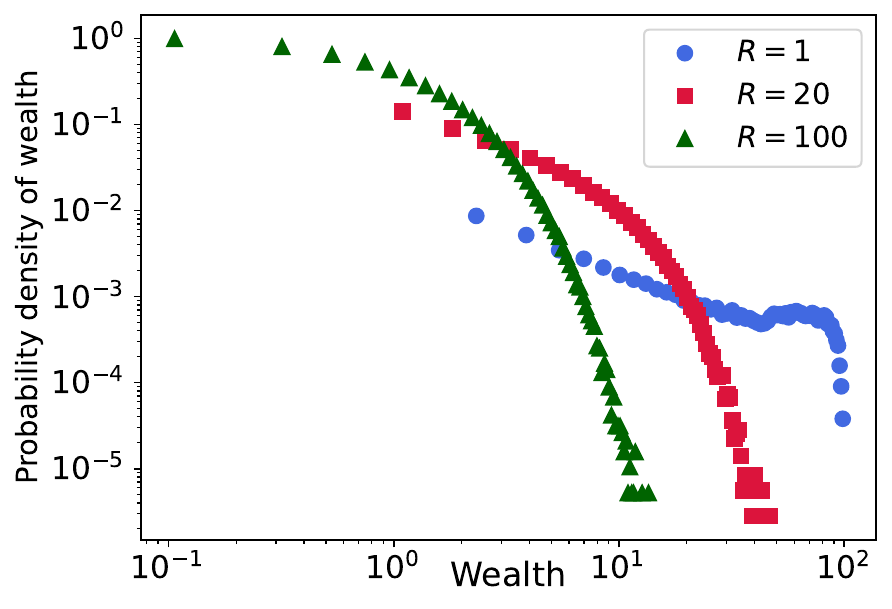}
\end{figure}
FIG. I: \textbf{B-Model:} Probability distribution functions with different Exchange Range $R$. From these distribution functions different inequality indices $g$, $p$ and $k$ have computed for different $R$ as reported in Table~\ref{tab:1}.
\begin{figure}[H]
    \centering    
    \includegraphics[width=0.48\linewidth]{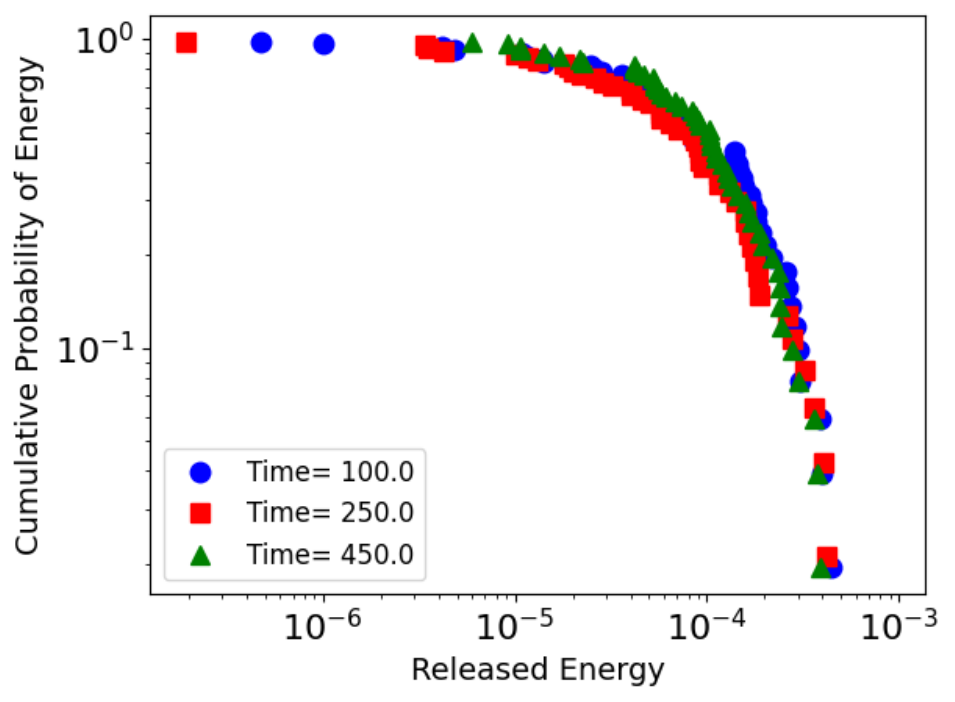}
\end{figure}
FIG. II: \textbf{BK-Model:} Cumulative probability distribution at different times. From these distribution functions the inequality indices $g$, $p$ and $k$ have calculated.

\section{B-Model with different Exchange Range $\mathbf{R}$, B-Model with different Savings $\mathbf{\lambda}$ and C-Model with different Savings $\mathbf{\lambda}$ for population $\mathbf{N=200}$}\label{app4}
\begin{figure}[H]
    \centering   
    \includegraphics[width=0.33\linewidth]{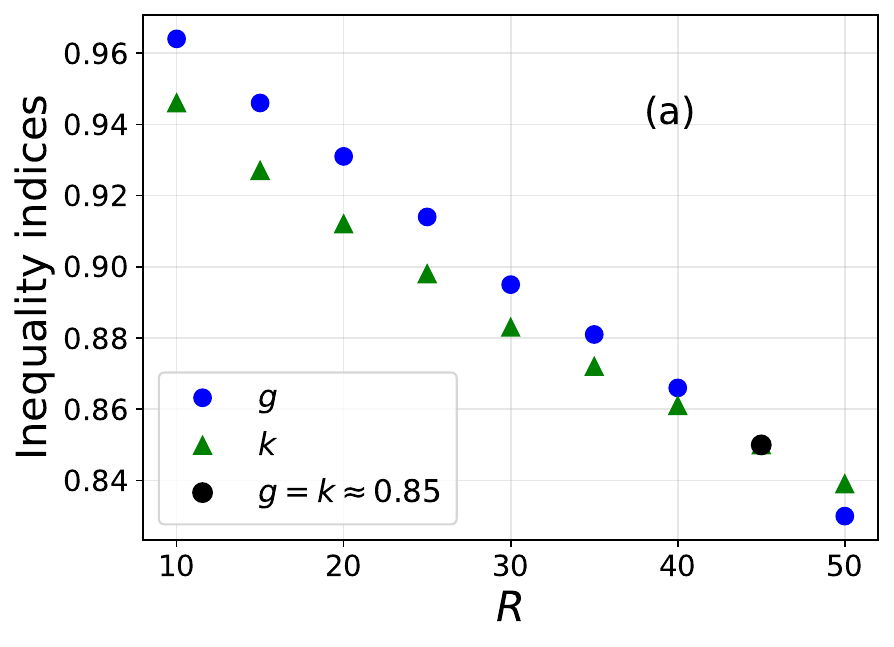}
    \includegraphics[width=0.33\linewidth]{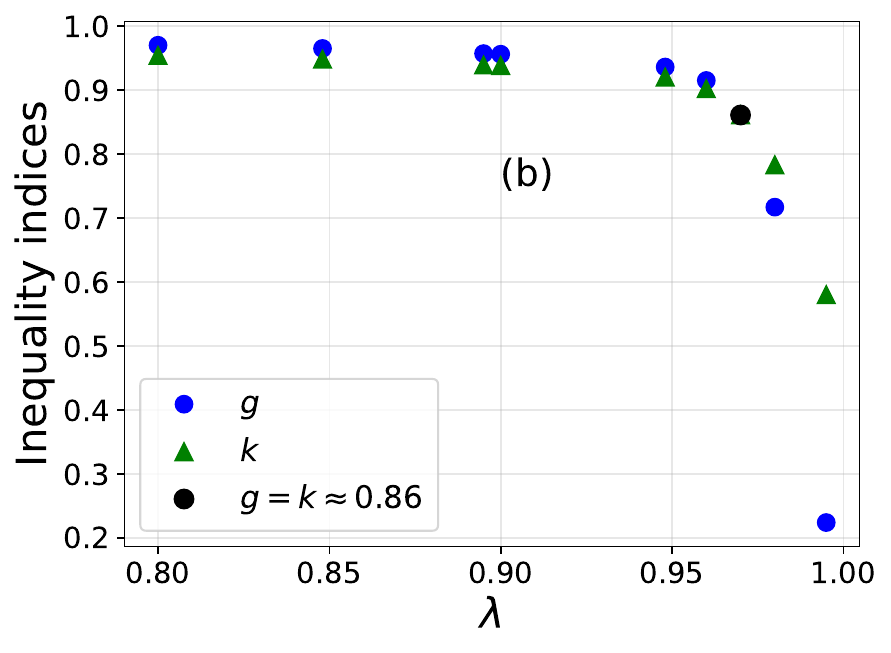}\includegraphics[width=0.33\linewidth]{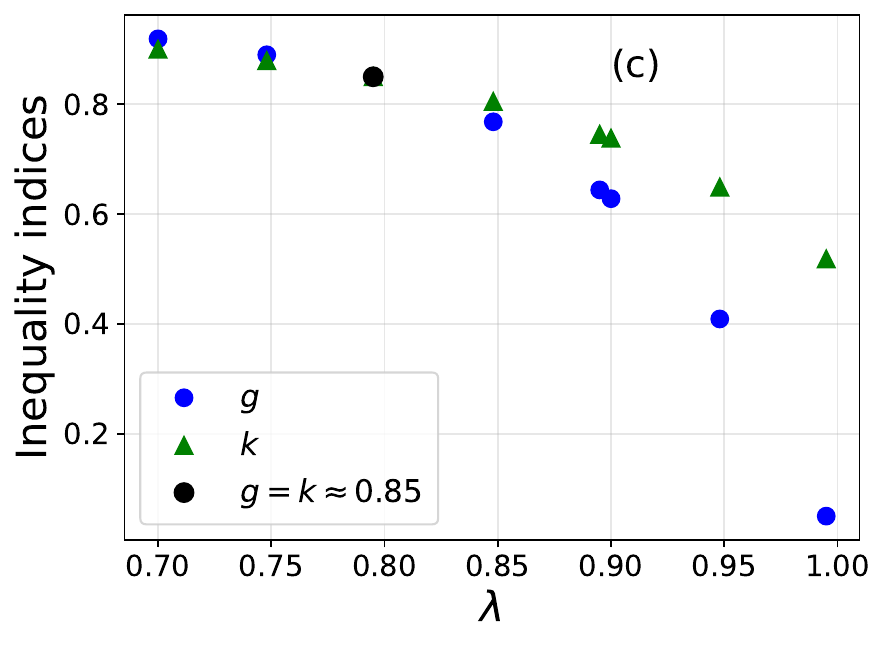}
\end{figure}
FIG. III: \textbf{B-Model with $R$ and $\lambda$ and C-Model with $\lambda$:} Numerical results for the variations of inequality indices $g$ and $k$ for total population $N=200$. Variations (a) with range $R$ for B-Model; (b) and (c) with saving propensity $\lambda$ in B-Model and C-Model respectively.

\bibliography{ref} 

@article{DY_2000,
  author = {Dragulescu, Adrian and Yakovenko, Victor M.},
  title = {Statistical Mechanics of Money},
  journal = {The European Physical Journal B-Condensed Matter and Complex Systems}, 
  volume={17}, 
  pages={723--729}, 
  year={2000},
  doi = {https://doi.org/10.1007/s100510070114}
}

@article{CC_2000,
  title={Statistical mechanics of money: how saving propensity affects its distribution},
  author={Chakraborti, Anirban and Chakrabarti, Bikas K},
  journal={The European Physical Journal B-Condensed Matter and Complex Systems},
  volume={17},
  number={1},
  pages={167--170},
  year={2000},
  publisher={Springer},
  doi = {https://doi.org/10.1007/s100510070173}
}

@article{CCM_2004,
  title={Pareto law in a kinetic model of market with random saving propensity},
  author={Chatterjee, Arnab and Chakrabarti, Bikas K and Manna, Subhrangshu Sekhar},
  journal={Physica A: Statistical Mechanics and its Applications},
  volume={335},
  number={1-2},
  pages={155--163},
  year={2004},
  publisher={Elsevier},
  doi = {https://doi.org/10.1016/j.physa.2003.11.014}
}

@article{C_Model,
  title={Distributions of money in model markets of economy},
  author={Chakraborti, Anirban},
  journal={International Journal of Modern Physics C},
  volume={13},
  number={10},
  pages={1315--1321},
  year={2002},
  publisher={World Scientific},
  doi = {https://doi.org/10.1142/S0129183102003905}
}

@article{YS,
  title={Is inequality inevitable},
  author={Boghosian, Bruce M},
  journal={Scientific American},
  volume={321},
  number={5},
  pages={70--77},
  year={2019},
  url = {https://www.scientificamerican.com/article/is-inequality-inevitable/}
}

@article{YS1,
  title = {Kinetics of wealth and the Pareto law},
  author = {Boghosian, Bruce M.},
  journal = {Physical Review E—Statistical, Nonlinear, and Soft Matter Physics},
  volume = {89},
  issue = {4},
  pages = {042804},
  numpages = {22},
  year = {2014},
  month = {Apr},
  publisher = {American Physical Society},
  doi = {https://doi.org/10.1103/PhysRevE.89.042804}
}

@ARTICLE{B_Model,
       author = {{Banerjee}, Suchismita},
        title = {Role of Neighbouring Wealth Preference in Kinetic Exchange model of market},
      journal = {arXiv e-prints},
         year = {2023},
          doi = {https://doi.org/10.48550/arXiv.2305.16238}
}

@Article{B_Model1,
AUTHOR = {Ghosh, Asim and Banerjee, Suchismita and Goswami, Sanchari and Mitra, Manipushpak and Chakrabarti, Bikas K.},
TITLE = {Kinetic Models of Wealth Distribution with Extreme Inequality: Numerical Study of Their Stability against Random Exchanges},
JOURNAL = {Entropy},
VOLUME = {25},
YEAR = {2023},
NUMBER = {7},
ARTICLE-NUMBER = {1105},
ISSN = {1099-4300},
DOI = {https://doi.org/10.3390/e25071105}
}

@book{Pareto,
    author = {Pareto, Vilfredo},
    title = {Cours d’économie politique},
    publisher = {Librairie Droz},
    year = {1964}
}

@article{Pareto1,
author = {Merritt, Fred D},
title = {Cours d'Economie Politique. Vilfredo Pareto },
journal = {Journal of Political Economy},
volume = {6},
number = {4},
pages = {549-552},
year = {1898},
doi = {https://doi.org/10.1086/250536}
}

@article{CS_Model,
title={Stick-slip statistics for two fractal surfaces: a model for earthquakes},
author={Chakrabarti, Bikas K and Stinchcombe, Robin B},
journal={Physica A: Statistical Mechanics and its Applications},
volume={270},
number={1-2},
pages={27--34},
year={1999},
publisher={Elsevier},
doi = {https://doi.org/10.1016/S0378-4371(99)00146-6}
}

@article{CS1,
  title = {Statistical physics of fracture, friction, and earthquakes},
  author = {Kawamura, Hikaru and Hatano, Takahiro and Kato, Naoyuki and Biswas, Soumyajyoti and Chakrabarti, Bikas K.},
  journal = {Review of Modern Physics},
  volume = {84},
  issue = {2},
  pages = {839--884},
  numpages = {0},
  year = {2012},
  month = {May},
  publisher = {American Physical Society},
  doi = {https://doi.org/10.1103/RevModPhys.84.839}
}

@article{BK_Model,
author = {Mori, Takahiro and Kawamura, Hikaru},
title = {Spatiotemporal correlations of earthquakes in the continuum limit of the one-dimensional Burridge-Knopoff model},
journal = {Journal of Geophysical Research: Solid Earth},
volume = {113},
number = {B11},
doi = {https://doi.org/10.1029/2008JB005725},
year = {2008}
}

@article{langer_1994,
    author = {Carlson, J. M. and Langer, J. S. and Shaw, B. E.},
    title = {Dynamics of earthquake faults},
    journal = {Review of Modern Physics},
    volume = {66},
    pages = {657},
    publisher = {American Physical Society},
    doi = {https://doi.org/10.1103/RevModPhys.66.657},
    year = {1994}
}

@article{lor,
  title={Methods of measuring the concentration of wealth},
  author={Lorenz, Max O},
  journal={Publications of the {A}merican {S}tatistical {A}ssociation},
  volume={9},
  number={70},
  pages={209--219},
  year={1905},
  publisher={Taylor \& Francis},
  doi = {https://doi.org/10.2307/2276207}
}

@article{lor1,
  title={Characterizations of Lorenz curves and income distributions},
  author={Aaberge, Rolf},
  journal={Social Choice and Welfare},
  volume={17},
  number={4},
  pages={639--653},
  year={2000},
  publisher={Springer},
  doi = {https://www.jstor.org/stable/41106382}
}

@book{gini,
author = {Corrado Gini},
title = {Variabilit{\`a} e Mutabilit{\`a}: Contributo allo Studio delle Distribuzioni e delle Relazioni Statistiche},
address = {Bologna},
publisher = {C. Cuppini},
year = {1912}
}

@book{pietra,
  title={Delle relazioni tra gli indici di variabilita: nota 1},
  author={Pietra, Gaetano},
  year={1915},
  publisher={Ferrari}
}

@article{pietra1,
  title={Measuring statistical heterogeneity: The Pietra index},
  author={Eliazar, Iddo I and Sokolov, Igor M},
  journal={Physica A: Statistical Mechanics and its Applications},
  volume={389},
  number={1},
  pages={117--125},
  year={2010},
  publisher={Elsevier},
  doi = {https://doi.org/10.1016/j.physa.2009.08.006}
}

@article{k-index,
  title={Inequality in societies, academic institutions and science journals: Gini and k-indices},
  author={Ghosh, Asim and Chattopadhyay, Nachiketa and Chakrabarti, Bikas K},
  journal={Physica A: Statistical Mechanics and its Applications},
  volume={410},
  pages={30--34},
  year={2014},
  publisher={Elsevier},
  doi = {https://doi.org/10.1016/j.physa.2014.05.026}
}

@article{k-index1,
  title={Socio-economic inequality: Relationship between Gini and Kolkata indices},
  author={Chatterjee, Arnab and Ghosh, Asim and Chakrabarti, Bikas K},
  journal={Physica A: Statistical Mechanics and its Applications},
  volume={466},
  pages={583--595},
  year={2017},
  publisher={Elsevier},
  doi = {https://doi.org/10.1016/j.physa.2016.09.027}
}

@article{k-index2,
  title={On the Kolkata index as a measure of income inequality},
  author={Banerjee, Suchismita and Chakrabarti, Bikas K and Mitra, Manipushpak and Mutuswami, Suresh},
  journal={Physica A: Statistical Mechanics and Its Applications},
  volume={545},
  pages={123178},
  year={2020},
  publisher={Elsevier},
  doi = {https://doi.org/10.1016/j.physa.2019.123178}
}

@article{k-index3,
  title={Inequality measures: The Kolkata index in comparison with other measures},
  author={Banerjee, Suchismita and Chakrabarti, Bikas K and Mitra, Manipushpak and Mutuswami, Suresh},
  journal={Frontiers in Physics},
  volume={8},
  pages={562182},
  year={2020},
  publisher={Frontiers Media SA},
  doi = {https://doi.org/10.3389/fphy.2020.562182}
}

@Article{sand,
AUTHOR = {Banerjee, Suchismita and Biswas, Soumyajyoti and Chakrabarti, Bikas K. and Ghosh, Asim and Mitra, Manipushpak},
TITLE = {Sandpile Universality in Social Inequality: Gini and Kolkata Measures},
JOURNAL = {Entropy},
VOLUME = {25},
YEAR = {2023},
NUMBER = {5},
ARTICLE-NUMBER = {735},
DOI = {https://doi.org/10.3390/e25050735}
}

@article{bijin,
  author = {Bijin Joseph and Bikas K. Chakrabarti},
  title = {Variation of Gini and Kolkata indices with saving propensity in the Kinetic Exchange model of wealth distribution: An analytical study},
  journal = {Physica A: Statistical Mechanics and its Applications},
  volume = {594},
  pages = {127051},
  year = {2022},
  doi = {https://doi.org/10.1016/j.physa.2022.127051}
}

@ARTICLE{asim_2025,
       author = {Ghosh, Asim and Chakrabarti, Bikas K},
        title = {Relations Between the Inequality Indices Gini, Pietra and Kolkata: Theory and Data Analysis},
      journal = {arXiv e-prints},
         year = {2025},
          doi = {https://doi.org/10.48550/arXiv.2512.00754}
}

@article{manna,
author = {S.S. Manna and Soumyajyoti Biswas and Bikas K. Chakrabarti},
title = {Near universal values of social inequality indices in self-organized critical models},
journal = {Physica A: Statistical Mechanics and its Applications},
volume = {596},
pages = {127121},
year = {2022},
issn = {0378-4371},
doi = {https://doi.org/10.1016/j.physa.2022.127121}
}

@article{soumya,
  title = {Critical Scaling through Gini Index},
  author = {Das, Soumyaditya and Biswas, Soumyajyoti},
  journal = {Physical Review Letters},
  volume = {131},
  issue = {15},
  pages = {157101},
  numpages = {6},
  year = {2023},
  month = {Oct},
  publisher = {American Physical Society},
  doi = {https://doi.org/10.1103/PhysRevLett.131.157101}
}

@article{pbhattacharya,
  title = {Of overlapping Cantor sets and earthquakes: analysis of the discrete Chakrabarti–Stinchcombe model},
  author = {Bhattacharya, Pratip},
  journal = {Physica A: Statistical Mechanics and its Applications},
  volume = {348},
  pages = {199-215},
  numpages = {6},
  year = {2005},
  month = {Mar},
  publisher = {Elsevier},
  doi = {https://doi.org/10.1016/j.physa.2004.09.014}
}

@article{unequalearthquakes_soumya,
    author = {{Sarkar}, Sudip and {Biswas}, Soumyajyoti},
    title = {Large earthquakes follow highly unequal ones},
  journal = {arXiv e-prints},
  year = {2026},
  doi = {10.48550/arXiv.2601.08356}
}

\end{document}